\documentclass[twocolumn]{aastex63}
\usepackage{newtxtext,newtxmath}
\usepackage{graphicx}

\shorttitle{An H$_{2}$-bearing DLA at $z=0.576$}
\shortauthors{Boettcher et al.}

\begin{document}

\title{The Cosmic Ultraviolet Baryon Survey (CUBS) II: Discovery of an H$_{2}$-Bearing DLA in the Vicinity of an Early-Type Galaxy at $z = 0.576$\footnote{Based on data gathered with the 6.5m Magellan Telescopes located at Las Campanas Observatory and the NASA/ESA Hubble Space Telescope operated by the Space Telescope Science Institute and the Association of Universities for Research in Astronomy, Inc., under NASA contract NAS 5-26555.}}

\correspondingauthor{Erin Boettcher}
\email{eboettcher@astro.uchicago.edu}

\author{Erin Boettcher}
\affiliation{Department of Astronomy \&
  Astrophysics, The University of Chicago, 5640 S. Ellis Ave.,
  Chicago, IL 60637, USA}

\author{Hsiao-Wen Chen}
\affiliation{Department of Astronomy \&
  Astrophysics, The University of Chicago, 5640 S. Ellis Ave.,
  Chicago, IL 60637, USA}

\author{Fakhri S. Zahedy}
\affiliation{The Observatories of the Carnegie Institution for Science, 813 Santa Barbara Street, Pasadena, CA 91101, USA}

\author{Thomas J. Cooper}
\affiliation{The Observatories of the Carnegie Institution for Science, 813 Santa Barbara Street, Pasadena, CA 91101, USA}

\author{Sean D. Johnson}
\affiliation{Department of Astronomy,
  University of Michigan, Ann Arbor, MI 48109, USA}

\author{Gwen C. Rudie}
\affiliation{The Observatories of the Carnegie Institution for Science, 813 Santa Barbara Street, Pasadena, CA 91101, USA}

\author{Mandy C. Chen}
\affiliation{Department of Astronomy \&
  Astrophysics, The University of Chicago, 5640 S. Ellis Ave.,
  Chicago, IL 60637, USA}

\author{Patrick Petitjean}
\affiliation{Institut d'Astrophysique de Paris, CNRS-SU, UMR 7095, 98bis bd Arago, Paris F-75014, France}

\author{Sebastiano Cantalupo}
\affiliation{Department of Physics, ETH Z\"{u}rich, Wolfgang-Pauli-Strasse 27, Z\"{u}rich, 8093 Switzerland}

\author{Kathy L. Cooksey}
\affiliation{Department of Physics and Astronomy, University of Hawai’i at Hilo, Hilo, HI 96720, USA}

\author{Claude-Andr\'{e} Faucher-Gigu\`{e}re}
\affiliation{Department of Physics \& Astronomy and Center for Interdisciplinary Exploration and Research in Astrophysics (CIERA), Northwestern University, 1800 Sherman Ave, Evanston, IL 60201, USA}

\author{Jenny E. Greene}
\affiliation{Department of Astrophysics, Princeton University, Princeton, NJ 08544, USA}

\author{Sebastian Lopez}
\affiliation{Departamento de Astronom\'{i}a, Universidad de Chile, Casilla 36-D, Santiago, Chile}

\author{John S. Mulchaey}
\affiliation{The Observatories of the Carnegie Institution for Science, 813 Santa Barbara Street, Pasadena, CA 91101, USA}

\author{Steven V. Penton}
\affiliation{Laboratory For Atmospheric and Space Physics, University of Colorado, Boulder, CO 80303, USA}

\author{Mary E. Putman}
\affiliation{Department of Astronomy, Columbia University, New York, NY 10027, USA}

\author{Marc Rafelski}
\affiliation{Space Telescope Science Institute, Baltimore, MD 21218, USA}
\affiliation{Department of Physics \& Astronomy, Johns Hopkins University, Baltimore, MD 21218, USA}

\author{Michael Rauch}
\affiliation{The Observatories of the Carnegie Institution for Science, 813 Santa Barbara Street, Pasadena, CA 91101, USA}

\author{Joop Schaye}
\affiliation{Leiden Observatory, Leiden University, PO Box 9513, NL-2300 RA Leiden, the Netherlands}

\author{Robert A. Simcoe}
\affiliation{MIT-Kavli Institute for Astrophysics and Space Research, 77 Massachusetts Ave., Cambridge, MA 02139, USA}

\author{Gregory L. Walth}
\affiliation{The Observatories of the Carnegie Institution for Science, 813 Santa Barbara Street, Pasadena, CA 91101, USA}



\begin{abstract}

We report the serendipitous detection of an H$_{2}$-bearing damped
Lyman-$\alpha$ absorber at $z = 0.576$ in the spectrum of the QSO
J$0111\!-\!0316$ in the Cosmic Ultraviolet Baryon
Survey. Spectroscopic observations from \textit{HST}-COS in the
far-ultraviolet reveal a damped absorber with
log[$N$(\ion{H}{1})/cm$^{-2}$] $= 20.1 \pm 0.2$ and
log[$N$(H$_{2}$)/cm$^{-2}$] $= 18.97\substack{+0.05 \\ -0.06}$.  The
diffuse molecular gas is found in two velocity components separated by
$\Delta\,\varv \approx 60$ km\,s$^{-1}$, with $>99.9$\% of the total
H$_{2}$ column density concentrated in one component. At a metallicity
of $\approx 50$\% of solar, there is evidence for Fe enhancement and
dust depletion, with a dust-to-gas ratio $\kappa_{\text{O}} \approx
0.4$. A galaxy redshift survey conducted with IMACS and LDSS-3C on
Magellan reveals an overdensity of nine galaxies at projected distance
$d \le 600$ proper kpc (pkpc) and line-of-sight velocity offset
$\Delta\,\varv_{g} \le 300$ km\,s$^{-1}$ from the absorber. The
closest is a massive, early-type galaxy at $d = 41$ pkpc which
contains $\approx 70$\% of the total stellar mass identified at $d \le
310$ pkpc of the H$_{2}$ absorber. The close proximity of the
H$_{2}$-bearing gas to the quiescent galaxy and the Fe-enhanced
chemical abundance pattern of the absorber suggest a physical
connection, in contrast to a picture in which DLAs are primarily
associated with gas-rich dwarfs. This case study illustrates that deep
galaxy redshift surveys are needed to gain insight into the diverse
environments that host dense and potentially star-forming gas.

\end{abstract}



\section{Introduction}
\label{sec:intro}

The origin and extent of diffuse molecular gas in the circumgalactic
medium (CGM) is an open question in astrophysics. The presence of
circumgalactic molecules has important implications for the diversity
of environments in which potentially star-forming gas is produced
and/or transported. The Lyman and Werner band electronic transitions
of H$_{2}$ ($\lambda_{\text{rest}} \sim 900 - 1100$ \AA) provide a
powerful tool for detecting and characterizing diffuse molecular gas
in absorption against bright background QSOs. The H$_{2}$-bearing
damped Lyman-$\alpha$ absorbers (with neutral hydrogen column density
log[$N$(\ion{H}{1})/cm$^{-2}$] $\gtrsim 20.3$, DLAs) and strong Lyman
limit systems ($19.0 \lesssim$~log[$N$(\ion{H}{1})/cm$^{-2}$]
$\lesssim 20.3$, sLLSs) probed by these transitions thus provide a
critical class of objects for characterizing the presence and
properties of molecular gas in a range of galactic environments.

At high redshift where the Lyman and Werner band transitions are
accessible from the ground ($z \gtrsim 1.8$), over 100 H$_{2}$-bearing
DLAs and sLLSs have been detected through systematic searches
\citep[e.g.,][]{Ledoux2003, Noterdaeme2008} and mining of data
archives, including the Sloan Digital Sky Survey and follow-up
observations \citep[e.g.,][]{Balashev2014, Balashev2018, Balashev2019,
  Noterdaeme2019}. At low redshift ($z \lesssim 1$), \textit{Hubble
  Space Telescope} (\textit{HST}) observations in the ultraviolet (UV)
have revealed less than 20 such systems \citep{Crighton2013,
  Oliveira2014, Srianand2014, Dutta2015, Muzahid2015, Muzahid2016,
  Zahedy2020}. There is evidence for a higher occurance rate of
molecules in DLAs and sLLSs with higher metallicities, dust-to-gas
ratios, and absorption line widths of low ionization states
\citep[e.g.,][]{Ledoux2003, Petitjean2006, Noterdaeme2008}. The
correlation between the incidence of H$_{2}$-bearing gas and the line
width may be due to star-formation feedback; alternatively, it may be
secondary evidence of a metallicity dependence due to a
mass-metallicity relationship \citep{Ledoux2006_mm}. These factors
suggest a picture in which H$_{2}$-bearing absorbers are associated
with metal-enriched, star-forming environments.

A key question in interpreting the nature of H$_{2}$-bearing DLAs and
sLLSs is whether they arise in the bound ISM of galaxies or are
evidence of a circumgalactic molecular medium. At low redshift, the
galactic environments of absorbers can be characterized in detail via
ground-based galaxy redshift surveys. Of the 12 H$_{2}$-bearing
absorbers detected beyond the Milky Way at $z < 1$, 10 have candidate
host galaxies, eight of which are spectroscopically confirmed to be at
the redshift of the absorber \citep{Crighton2013, Oliveira2014,
  Srianand2014, Muzahid2015, Muzahid2016, Zahedy2020}. Eight of these
host galaxy candidates have impact parameters $10 \lesssim d \lesssim
80$ proper kpc (pkpc).

If these galaxies are indeed the hosts of the absorbing gas, this
implies a picture in which H$_{2}$-bearing DLAs and sLLSs arise
primarily outside of the inner disks of the identified
galaxies. Suggested origins for the diffuse molecular gas include
faint, undetected satellites, outflows from star-forming galaxies, and
gas stripped from galaxies or their companions by tidal forces or ram
pressure \citep[e.g.,][]{Dutta2015, Muzahid2015}. Galactic outflows
from star-forming satellites may also contribute significantly to cool
gas in the halos of massive galaxies \citep[e.g.,][]{Hafen2019,
  FaucherGiguere2016}. However, our understanding of the host galaxies
of molecular absorbers is limited by the small number of known systems
and the absence of information about the large-scale galactic
environments of these absorbers. The association between an absorber
and a host galaxy candidate is ambiguous when additional galaxies
without known redshifts remain in the field, including the possible
presence of undetected dwarfs. It is also very difficult to detect
galaxies at small impact parameters due to the challenges of
subtracting the bright QSO light. The need remains for deep
photometric surveys and follow-up spectroscopic observations to
sub-$L_{\text{star}}$ sensitivities in the fields of H$_{2}$-bearing
DLAs and sLLSs to confidently characterize the galactic environments
and origins of these absorbers \citep{Muzahid2015}.

Here, we report the serendipitous detection of an H$_{2}$-bearing DLA
at $z = 0.576$ towards the background QSO J$0111\!-\!0316$ (RA,
decl. $=$ 01h11m39.170s, -03d16m10.89s; \citealt{Gaia2018}) in the
Cosmic Ultraviolet Baryon Survey (CUBS; \citealt{Chen2020}). At a
redshift of $z_{\text{QSO}} = 1.234$, the QSO (FUV = 18.5, NUV = 16.7;
\citealt{Bianchi2014}) is part of an NUV-bright sample studied with
the Cosmic Origins Spectrograph (COS) on \textit{HST} and an
accompanying ground-based galaxy redshift survey carried out
spectroscopically on Magellan (IMACS, LDSS-3C) and the Very Large
Telescope (MUSE). These observations permit a detailed
characterization of the physical conditions in the H$_{2}$-bearing gas
and its galactic environment.

The outline of this paper is as follows. In \S\ref{sec:data}, we
describe the space- and ground-based FUV and optical absorption-line
spectroscopy of J$0111\!-\!0316$, and we discuss the galaxy redshift
survey carried out in the field of the QSO. The following Section
details our absorption-line profile analysis. In \S\ref{sec:results},
we discuss the detection and characterization of H$_{2}$-bearing
gas, as well as the kinematics and chemical abundance pattern of the
accompaning neutral and ionized atomic gas. The following Section
describes the galaxy overdensity at the redshift of the
H$_{2}$-bearing DLA, and we consider possible origin scenarios for the
absorber in \S\ref{sec:diss}. We summarize and conclude in
\S\ref{sec:conc}. Throughout this work, we adopt the solar abundance
pattern of \cite{Asplund2009} and assume a $\Lambda$ cosmology with
$\Omega_{\text{M}} = 0.3$, $\Omega_{\Lambda} = 0.7$, and $H_{0} = 70$
km\,s$^{-1}$ Mpc$^{-1}$.

\section{Data}
\label{sec:data}

The CUBS survey design is described in detail in \citet{Chen2020}. We
summarize the relevant aspects of the ground- and space-based data
collection and reduction procedures here.

\subsection{\textit{HST}-COS observations}
\label{sec:HST}

We used \textit{HST}-COS to obtain FUV spectroscopy of
J$0111\!-\!0316$ between UT 4 - 29 July 2018 (Program ID: 15163; PI:
Chen). The total exposure time was $t_{\text{exp}} = 15148$ s and
$t_{\text{exp}} = 19710$ s for the G130M and G160M gratings,
respectively. We used two central wavelengths (cenwaves) in the G130M
regime and four in the G160M regime to provide continuous wavelength
coverage across the CCD detector gap. At each cenwave, we used either
two or four FP-POS settings to reduce the impact of fixed-pattern
noise and to prevent spectral gaps due to Lyman-$\alpha$-induced
detector ``holes.''

The wavelength coverage ranges from $1070 - 1790$ \AA, with a full
width at half maximum (FWHM) of the line spread function of 0.11
\AA~($\approx 21$ km\,s$^{-1}$) at 1600 \AA. The spectral resolution
element (resel) is sampled by $\approx 8$ pixels. We used a custom
pipeline to stack the \textbf{x1d} spectra output by
\texttt{CalCOS}\footnote{https://github.com/spacetelescope/calcos/}
(v3.3.5) for each grating, cenwave, and FP-POS and perform the
continuum normalization \citep{Chen2018}. For the G160M grating, which
is the spectral region of interest for the $z = 0.576$ DLA reported
here, the signal-to-noise ratio (S/N) per spectral resolution element
decreases from S/N $\approx 27$ resel$^{-1}$ at the blue end to S/N
$\approx 15$ resel$^{-1}$ at the red end, falling below S/N $= 15$
resel$^{-1}$ only redward of 1750 \AA.

\subsection{MIKE observations}
\label{sec:MIKE}

We used the Magellan Inamori Kyocera Echelle (MIKE;
\citealt{Bernstein2003}) spectrograph on the Magellan Clay telescope
to obtain optical spectroscopy of J$0111\!-\!0316$ on UT 1 Nov
2018. The blue (red) arm provides spectral coverage from $3335 - 5060$
\AA\ ($4850 - 9400$ \AA). With a $0.7''$ slit width, the instrumental
line spread function has a FWHM of 7.5 km\,s$^{-1}$ in the blue arm
and 10 km\,s$^{-1}$ in the red. There are 2.1 (2.0) pixels sampling a
spectral resolution element in the blue (red) arm using 2 x 2 on-chip
binning. An exposure time of $t_{\text{exp}} = 3 \times 600$ s
produced S/N $\leq 60$ resel$^{-1}$ in the blue and S/N $\leq 80$
resel$^{-1}$ in the red; this drops to S/N $\approx 15$ resel$^{-1}$
at the ends of the useful spectral range. \citet{Chen2014} and
\citet{Zahedy2016} discuss the spectral reduction, extraction,
relative flux calibration, and continuum normalization using custom
software. In addition to enabling access to an expanded suite of ionic
transitions, the MIKE spectroscopy has improved spectral resolution
with respect to COS by a factor of $2 - 3$. These observations thus
serve as an independent verification of the COS wavelength calibration
and provide guidance for the kinematic analysis of COS absorbers (see
\S\ref{sec:analysis}).

\subsection{Magellan galaxy spectroscopy}
\label{sec:mag}

The Magellan galaxy survey design consists of two components: (1) a
{\it deep and narrow} component with Magellan/LDSS-3C that targets all
galaxies with $L > 0.1~L_{\text{star}}$ and $z \lesssim 1$ (and fainter
at lower redshifts) within $\lesssim 1'$ in angular radius from the
QSO sightline ($d < 325\!-\!500$ pkpc for $0.4<z<1.0$) and (2) a
{\it shallow and wide} component with Magellan/IMACS \citep{IMACS}
that targets $L_{\text{star}}$ galaxies at $\lesssim 10'$ ($d<3-5$
pMpc) from the QSO sightline.

For the {\it deep and narrow} LDSS-3C survey, we aim to establish a
statistically representative sample of galaxies at $z<1$ for a
comprehensive study of the CGM on individual halo scales ($d<300$ pkpc)
across a broad range in both stellar mass and star formation rate.  To
increase the efficiency of our ground-based survey, we devised a
color-based prioritization scheme using the Ultra-VISTA survey
\citep{Muzzin2013}. We prioritize galaxies with $r'< 24$ mag within
the inner $1'$ surrounding the QSO sightline with colors satisfying
$r-H < 2\times (g-r)+0.6$ and $r-H<3.2$. These criteria select nearly
all log$(M_{\text{star}}/\text{M}_{\odot}) > 10$ galaxies (both
star-forming and passive) at $z<0.8$ and 50\% at $z=0.8-1$.

We obtained LDSS-3C data on UT 4 - 6 Oct 2018 and UT 2 - 3 Nov 2018
using the VPH-all 400 l/mm grism and a $1''$ slit width, yielding
wavelength coverage from $4250 - 10000$ \AA\ at a spectral resolution
of $R = 650$ or $\varv_{\text{FWHM}} = 460$ km\,s$^{-1}$. The exposure
time was $t_{\text{exp}} = 6000 - 10800$ s for each of the five masks
probing $\sim 30$ galaxies each, with more exposure time alloted to
masks probing fainter galaxies. On UT 8 and 9 Sept and UT 9 Nov 2018,
we obtained IMACS observations using the f/2 camera, 150 l/mm grism,
and a $1''$ slit width, yielding wavelength coverage from $5000 -
9000$ \AA\ with a spectral resolution of $R = 550$
($\varv_{\text{FWHM}} = 550$ km\,s$^{-1}$) at the blaze wavelength of
$7200$ \AA. The observations were divided among four masks, with $\sim
300$ galaxies observed for $t_{\text{exp}} = 5400 - 6300$ s per mask.

Multislit spectral reductions were performed using the CPDMap routine
in the
\texttt{CarPy}\footnote{https://code.obs.carnegiescience.edu/carnegie-python-distribution}
Python distribution, which implements techniques for rectification and
sky subtraction as outlined in \citet{Kelson2000} and
\citet{Kelson2003}. Redshifts were measured using custom software that
constructs a template spectrum at regular redshift intervals by
performing a $\chi^{2}$ minimization of a linear combination of the
galaxy eigenspectra from \citet{Bolton2012}. The redshift with the
overall minimum $\chi^{2}$ was then selected as the galaxy redshift,
with manual verification and exclusion of poor spectral regions.

\section{Analysis}
\label{sec:analysis}

We characterize the column density ($N$), velocity ($\varv$), and
Doppler parameter ($b$) of all absorbers detected in the COS and MIKE
spectroscopy through a Voigt profile analysis. The transition
wavelengths and oscillator strengths are taken from \texttt{vpfit}
(v11.1) \citep{Carswell2014} for the atomic lines and
\citet{Ubachs2019} for the molecular lines. Using custom software, we
fit the absorption lines for a given atomic, ionic, or molecular
species with a Voigt profile function by minimizing the $\chi^{2}$
statistic on a grid in log($N$), $\varv$, $b$ parameter space. We define $\chi^{2}$ as usual,
\begin{equation}
    \chi^{2} = \sum_{i=0}^{n} \frac{(f_{\text{obs},i} - f_{\text{mod},i})^{2}}{\sigma_{\text{obs},i}^{2}},
	\label{eq:quadratic}
\end{equation}
where $f_{\text{obs},i}$ and $f_{\text{mod},i}$ are the observed and
modeled flux per spectral pixel and $\sigma_{\text{obs},i}$ is the
associated flux uncertainty. The summation is over the wavelength
range that encompasses the extent of the absorption, excluding
contaminated regions.

We take the minimum $\chi^{2}$ value as an estimate of the best-fit
model parameters; we then refine these best-fit values and calculate
their corresponding uncertainties using a Markov Chain Monte Carlo
(MCMC) method that employs the Metropolis-Hastings algorithm
\citep[e.g.,][]{Ivezic2014}. Here, the likelihood of a given model
follows from the $\chi^{2}$ statistic as $\mathcal{L} \propto
\mathrm{e}^{-\chi^{2}/2}$, and we impose model priors as discussed
below. In summary, we apply no priors to fits of the suite of
\ion{Fe}{2} lines available in the MIKE data, and we require the
remainder of the low ions observed with MIKE and COS to have the same
$\varv$ and $b$ values for matching velocity components, where the COS
lines are allowed to vary within wavelength calibration uncertainties
of $\delta\,\varv = \pm 5$ km\,s$^{-1}$. Priors imposed on the
\ion{H}{1} and H$_{2}$ absorption-line analyses are discused in
\S\ref{sec:atom} and \S\ref{sec:H2}, respectively.

We begin each MCMC run near the location of minimum $\chi^{2}$ in
parameter space for efficiency \citep[e.g.,][]{Zahedy2020}, run each
chain for $(2.5 - 10) \times 10^{4}$ steps, and discard the first 10\%
of the chain as the burn-in. The best-fit parameters correspond to the
50th percentile of the resultant marginalized posterior probability
distributions, and the reported uncertainties correspond to the 16th
and 84th percentiles of these same distributions. For non-detections,
we report an upper limit on the column density determined from a
$3\sigma$ upper limit on the equivalent width assuming a Doppler
parameter chosen to match other atomic or molecular transitions with
similar ionization or excitation states. This measurement is perfomed
on the strongest available non-contaminated transition. The
uncertainty analysis does not include possible modest contributions
from the choice of continuum fit and line spread function, except
where noted.

We leverage the high spectral resolution and S/N of the MIKE
observations to guide the absorption-line profile analysis. From
independent fits to the \ion{Mg}{2}\,$\lambda\lambda$2796, 2803 and
\ion{Fe}{2}\,$\lambda$2344, 2374, 2382, 2586, and 2600 transitions, we
determine that the line widths are dominated by turbulent rather than
thermal motions, and thus we adopt the same $b$ parameter across low
ionization states of different elements for a given velocity
component. We first determine the best-fit log($N$), $\varv$, and $b$
for the 10 components required to reproduce the absorption-line
profiles for the five \ion{Fe}{2} transitions; all other molecular,
atomic, and ionic transitions require equal or fewer components (see
Fig.~\ref{fig:HI_H2_metals}). Due to the wide range of oscillator
strengths among these five transitions (e.g.,
$f_{\lambda2374}/f_{\lambda2382} \approx 0.1$), there are always
multiple unsaturated transitions available. We then fix the values of
$\varv$ and $b$ for each component when fitting the remaining low ion
transitions available in MIKE and COS. Due to uncertainty in the COS
wavelength calibration on the order of a few km\,s$^{-1}$, we allow
the velocity centroid for all COS lines to vary around the best-fit
value from MIKE within $\delta\,\varv = \pm 5$ km\,s$^{-1}$.

For comparison, we also fit all metal-line profiles present in MIKE
and COS allowing $\varv$ and $b$ to vary between elements and
ionization states for a given component. The two approaches produce
comparable column densities, which we corroborate with a
curve-of-growth analysis. They also lead to consistent conclusions
about the kinematic state of the gas. We report the results from the
first fitting approach here. The saturation of the Lyman series lines
limits our characterization of the \ion{H}{1} velocity structure, and thus we
fit these transitions separately from those discussed above (see
\S~\ref{sec:atom}).

\section{Results}
\label{sec:results}

We detect absorption from the Lyman and Werner bands of molecular
hydrogen (H$_{2}$) at a redshift of $z = 0.576$ in the COS spectrum of
J$0111\!-\!0316$. The COS and MIKE data reveal absorption from
accompanying atomic hydrogen and a suite of neutral and ionized
metals. In this Section, we describe the column densities and
kinematics of the atomic (\S\ref{sec:atom}) and H$_{2}$-bearing gas,
including the excitation state of the molecular hydrogen
(\S\ref{sec:H2}). We also examine the chemical abundance pattern
(\S\ref{sec:chem_abundance}) and galactic environment
(\S\ref{sec:group}) of the absorber.

\subsection{Atomic gas properties}
\label{sec:atom}

In the COS spectrum, we detect absorption from the full Lyman series,
with the exception of the Ly$\alpha$ line which falls redward of our
wavelength coverage for the $z = 0.576$ absorber. Lyman continuum
photons are not transmitted by this damped absorber, and thus no
photons are detected blueward of an observed wavelength of $\approx 1445$
\AA.

\begin{figure*}
  \centering
  \includegraphics[scale = 0.9]{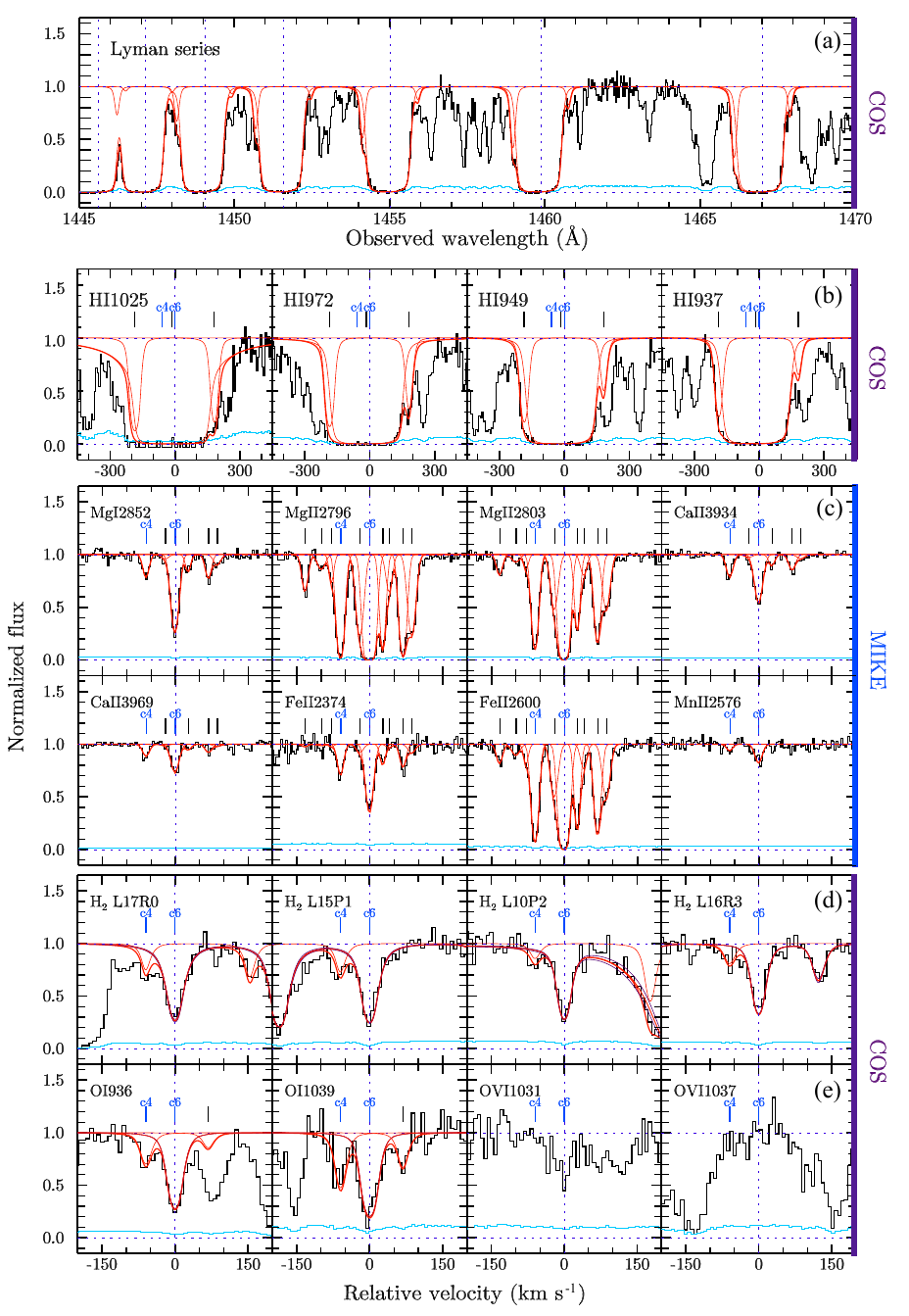}
  \caption{Observed (black) and modeled (red) line profiles of select
    \ion{H}{1} \textbf{(a, b)}, H$_{2}$ \textbf{(d)}, and metal
    transitions \textbf{(c, e)} from COS and MIKE spectra of the $z =
    0.576$ absorber towards J$0111\!-\!0316$. The COS spectrum is
    rebinned to three pixels, or $\Delta\,\varv \approx 8$
    km\,s$^{-1}$. The error spectra are shown in cyan. Individual
    velocity components are indicated with black tickmarks and shown
    as thin red lines; the cumulative model is indicated with a thick
    red line. The range of models within the reported uncertainties on
    $N$ and $b$ are shown by the thin purple lines for c6 in panels
    \textbf{(d)} and \textbf{(e)}. The H$_{2}$-bearing components, c4
    and c6, are labeled in blue in panels \textbf{(b)} -
    \textbf{(e)}. The dashed blue lines are shown to guide the eye at
    normalized flux values of zero and unity and at $\Delta\,\varv =
    0$ km\,s$^{-1}$, where the velocity zeropoint is set by the
    strongest metal line absorber at $z = 0.57616$. As shown in panel
    \textbf{(e)}, the absence of detected \ion{O}{6}\,$\lambda$1037
    absorption (log[$N$(\ion{O}{6})/cm$^{-2}$] < 14.0) suggests that
    the feature at \ion{O}{6}\,$\lambda$1031 is an interloper.}
  \label{fig:HI_H2_metals}
\end{figure*}

As listed in Table \ref{tab:COS_H} and shown in
Fig.~\ref{fig:HI_H2_metals}, a minimum of three velocity components
are required to reproduce the \ion{H}{1} absorption-line profiles (h1
- h3). The primary component has a column density of
log[$N$(\ion{H}{1})/cm$^{-2}$] $= 20.1 \pm 0.2$, where the uncertainty
accounts for both statistical and systematic error associated with the
continuum fitting of the Ly$\beta$ wings. Photoionization modeling
suggests that the \ion{H}{1} absorption arises from gas with a very
low ionization fraction (see \S\ref{sec:H2_ex}), and thus we refer to
this system as a DLA due to its largely neutral nature. The large $b$
value measured for this absorber ($b = 49$ km\,s$^{-1}$) likely
indicates the presence of multiple, unresolved components. The
velocity centroid is offset from the velocity zeropoint set by the
strongest metal-line absorber by $\Delta\,\varv = -16$
km\,s$^{-1}$. Components h1 and h3 contribute negligibly to the total
column density and do not have detectable associated metal lines, but
they indicate the presence of \ion{H}{1}-bearing gas over a velocity
range of $-190$ km\,s$^{-1}$ $\lesssim \Delta\,\varv \lesssim 180$
km\,s$^{-1}$.

\begin{deluxetable}{lccc}
  \label{tab:COS_H}
  \tablecaption{Atomic Hydrogen Properties}
  \tablehead{
    \colhead{} &
    \colhead{$\Delta\,\varv$} &
    \colhead{$b$} &
    \colhead{} \\
    \colhead{Comp.} &
    \colhead{(km\,s$^{-1}$)} &
    \colhead{(km\,s$^{-1}$)} &
    \colhead{log[$N$/cm$^{-2}$]}
  }
  \startdata
  h1 & $-186.3 \pm 0.6$ & $13.3\substack{+0.8 \\ -0.7}$ & $15.83 \pm 0.02$ \\
  h2 & $-15.7 \pm 0.3$ & $49.0 \pm 0.4$ & $20.1 \pm 0.2$ \\
  h3 & $180.9 \pm 0.7$ & $11.9 \pm 0.8$ & $14.97 \pm 0.04$ \\
  \noalign{\smallskip}
  \enddata
\end{deluxetable}

We detect absorption from neutral and ionized metals associated with
the DLA in both the COS and MIKE spectroscopy, including a suite of
\ion{O}{1} lines, \ion{O}{6}\,$\lambda\lambda$1031, 1037,
\ion{Mg}{1}\,$\lambda$2852, \ion{Mg}{2}\,$\lambda\lambda$2796, 2803,
\ion{Ca}{2}\,$\lambda\lambda$3934, 3969, \ion{Mn}{2}\,$\lambda$2576,
2594, 2606, and \ion{Fe}{2}\,$\lambda$2344, 2374, 2382, 2586, and
2600. As shown in Fig.~\ref{fig:HI_H2_metals}, the \ion{Fe}{2}
transitions require a minimum of 10 velocity components to reproduce
the absorption-line profiles at the high spectral resolution of the
MIKE data (c1 - c10); for all other atomic species, between two and 10
of these components show statistically significant absorption. The
velocity components range from $-130\ \text{km\,s$^{-1}$} \lesssim
\Delta\,\varv \lesssim 90$ km\,s$^{-1}$ and have a
component-to-component velocity dispersion of $\sigma_{\text{comp}}
\approx 70$ km\,s$^{-1}$. In Table \ref{tab:metals}, we summarize the
best-fit values of log($N$), $\Delta\,\varv$, and $b$, where the
velocity zeropoint is given by the strongest metal-line absorber in
MIKE, $z = 0.57616 \pm 0.00001$. The median best-fit $b$ value, $b
\approx 7.5$ km\,s$^{-1}$, significantly exceeds the thermal $b$ value
for a $T \lesssim 10^{4}$ K gas of $b_{\text{th}} \lesssim 2$
km\,s$^{-1}$, suggesting that the line width originates primarily in
turbulent rather than thermal motions.

\begin{deluxetable}{lcccc}
  \label{tab:metals}
  \tablecaption{Metal Line Properties}
  \tablehead{
    \colhead{} &
    \colhead{$\Delta\,\varv$} &
    \colhead{$b$} &
    \colhead{} &
    \colhead{} \\
    \colhead{Comp.} &
    \colhead{(km\,s$^{-1}$)} &
    \colhead{(km\,s$^{-1}$)} &
    \colhead{Ion} &
    \colhead{log[$N$/cm$^{-2}$]}
  }
  \startdata
  c1 & $-132.7 \pm 0.7$ & $7 \pm 1$ & \ion{Fe}{2} & $12.28 \pm 0.04$ \\
  & & & \ion{Mg}{2} & $12.17 \pm 0.02$ \\
  \noalign{\smallskip} 
  c2 & $-99 \pm 1$ & $10\substack{+3 \\ -2}$ & \ion{Fe}{2} & $12.13 \pm 0.08$ \\
  & & & \ion{Mg}{2} & $11.82\substack{+0.04 \\ -0.05}$ \\
  \noalign{\smallskip} 
  c3 & $-78.1 \pm 0.6$ & $5 \pm 1$ & \ion{Fe}{2} & $12.18 \pm 0.05$ \\
  & & & \ion{Mg}{2} & $11.91 \pm 0.03$ \\
  \noalign{\smallskip}
  c4 & $-59.6 \pm 0.1$ & $7.4 \pm 0.2$ & \ion{Fe}{2} & $13.44 \pm 0.01$ \\
  & & & \ion{Mg}{2} & $13.23 \pm 0.01$ \\
  & & & \ion{Mg}{1} & $11.44 \pm 0.04$ \\
  & & & \ion{Ca}{2} & $11.81 \pm 0.03$ \\
  & & & \ion{Mn}{2} & $11.6 \pm 0.1$ \\
  & & & \ion{O}{1} & $15.22 \pm 0.08$ \\
  \noalign{\smallskip}
  c5 & $-20.0\substack{+0.6 \\ -0.5}$ & $8.9\substack{+0.7 \\ -0.6}$ & \ion{Fe}{2} & $12.88 \pm 0.03$ \\
  & & & \ion{Mg}{2} & $12.77 \pm 0.01$ \\
  & & & \ion{Mg}{1} & $10.8\substack{+0.2 \\ -0.3}$ \\
  & & & \ion{Ca}{2} & $10.9\substack{+0.2 \\ -0.4}$ \\
  \noalign{\smallskip}
  c6 & $0.0 \pm 0.1$ & $9.2 \pm 0.2$ & \ion{Fe}{2} & $13.99 \pm 0.01$ \\
  & & & \ion{Mg}{2} & $> 14.03$\tablenotemark{a} \\
  & & & \ion{Mg}{1} & $12.26 \pm 0.01$ \\
  & & & \ion{Ca}{2} & $12.28 \pm 0.01$ \\
  & & & \ion{Mn}{2} & $12.16\substack{+0.03 \\ -0.04}$ \\
  & & & \ion{O}{1} & $16.48 \pm 0.06$ \\
  & & & \ion{O}{6} & $< 14.0$\tablenotemark{b} \\
  \noalign{\smallskip}
  c7 & $+27.1\substack{+0.2 \\ -0.1}$ & $5.8\substack{+0.3 \\ -0.2}$ & \ion{Fe}{2} & $13.16 \pm 0.01$ \\
  & & & \ion{Mg}{2} & $12.89 \pm 0.01$ \\
  & & & \ion{Mg}{1} & $11.23 \pm 0.06$ \\
  & & & \ion{Ca}{2} & $11.42 \pm 0.06$ \\
  \noalign{\smallskip}
  c8 & $+40.3 \pm 0.6$ & $7 \pm 2$ & \ion{Fe}{2} & $12.40 \pm 0.06$ \\
  & & & \ion{Mg}{2} & $12.18 \pm 0.02$ \\
  \noalign{\smallskip}
  c9 & $+68.6 \pm 0.1$ & $7.3 \pm 0.2$ & \ion{Fe}{2} & $13.28 \pm 0.01$  \\
  & & & \ion{Mg}{2} & $13.11 \pm 0.01$ \\
  & & & \ion{Mg}{1} & $11.45 \pm 0.04$ \\
  & & & \ion{Ca}{2} & $11.62 \pm 0.04$ \\
  & & & \ion{O}{1} & $14.8 \pm 0.1$ \\
  \noalign{\smallskip}
  c10 & $+86.8 \pm 0.3$ & $9.2 \pm 0.5$ & \ion{Fe}{2} & $12.91 \pm 0.02$ \\
  & & & \ion{Mg}{2} & $12.78 \pm 0.01$ \\
  & & & \ion{Mg}{1} & $11.14\substack{+0.07 \\ -0.08}$ \\
  & & & \ion{Ca}{2} & $11.1\substack{+0.1 \\ -0.2}$ \\
  \noalign{\smallskip}
  \enddata
  \tablenotetext{a}{Due to the saturation of \ion{Mg}{2} in c6, we report a 3$\sigma$ lower limit on $N$(\ion{Mg}{2}).}
  \tablenotetext{b}{The 3$\sigma$ upper limit on the \ion{O}{6} column density assuming a Doppler parameter of $b = 50$ km\,s$^{-1}$.}
\end{deluxetable}

\subsection{Diffuse molecular gas properties}
\label{sec:H2}

We detect absorption from molecular hydrogen associated with the
metal-line components c4 and c6. This absorption manifests as Lyman
and Werner band transitions that arise from the electronic ground
state with lower rotational levels $0 \leq J \leq 4$. The vibrational
level of the lower state is $v = 0$, while the upper state has $v \leq
18$ and $v \leq 5$ for the Lyman and Werner bands, respectively. This
produces approximately 230 observed absorption features between the
Lyman limit and the red cutoff of our spectral coverage, a significant
fraction of which are blended with each other or nearby atomic
transitions.

To characterize the properties of the H$_{2}$-bearing gas, we perform
Voigt profile fitting for the stronger (associated with c6,
$\Delta\,\varv = 0$ km\,s$^{-1}$) and weaker (c4, $\Delta\,\varv =
-59.6$ km\,s$^{-1}$) components separately. These absorbers are
sufficiently separated in velocity space to be fit individually with
the appropriate choice of mask. The fitting window is $-35$
km\,s$^{-1} \leq \varv \leq +45$ km\,s$^{-1}$ relative to the line
center for the c6 component; the asymmetry is due to the presence of
the c4 component. For the same reason, we choose a window of $\pm 20$
km\,s$^{-1}$ for the weaker c4 component, with the best-fit model for
the c6 component included in the fitting.

For each component, we fit transitions of a given $J$ value
simultaneously to determine log[$N_{J}$/cm$^{-2}$] and $b$ as
functions of $J$. Despite the large number of available transitions,
we find that the most robust results are obtained by fitting a small
number of carefully selected lines that are chosen due to their
relatively high S/N, lack of blending, and well-determined local
continuum (for example, we fit $J = 0$ for c6 based on the L0R0, L1R0,
and L17R0 transitions). In practice, we include between two
and six (one and three) transitions per $J$ value for the c6 (c4)
component. As for the COS metal lines, we allow the velocity centroid
of each transition to vary within $\pm 5$ km\,s$^{-1}$ of the best-fit
centroid determined from the MIKE metal lines.

We first performed the Voigt profile fitting for each $J$ level
separately, imposing the prior that $b \leq 10$ km\,s$^{-1}$ based on
the assumption that the turbulent line width traced by the molecules
does not exceed that given by the low ions. For the c6 component, the
model fitting for $J = 0$, $1$, and $4$ produces a single peak in the
probability distribution functions. However, for $J = 2$ and $3$,
there are two peaks at high (log[$N_{J}$/cm$^{-2}$] $\approx 18.0$)
and low (log[$N_{J}$/cm$^{-2}$] $\approx 16.0 - 16.5$) column density,
with low ($b \approx 4$ km\,s$^{-1}$) and high ($b \approx 7$
km\,s$^{-1}$) Doppler parameters, respectively. The two peaks are
well-separated in $N_{J}$, $b$ space for $J = 3$ but blended for $J =
2$.

For both $J$ levels, the most probable model corresponds to the high
column density solution. However, we know this solution to be
unphysical due to the absence of sufficiently high column density in
the $J = 4$ state (see Fig.~\ref{fig:temp}). The best-fit $b$ values
for $J = 0$ and $1$ ($b \approx 5$ km\,s$^{-1}$) are a factor of two
smaller than the best-fit value for $J = 4$ ($b = 9$ km\,s$^{-1}$),
suggesting that the Doppler parameter may rise with $J$ level (see
\S~\ref{sec:H2_ex} for a discussion of the physical origin(s) of this
trend). Additionally, as the $J \geq 2$ levels are largely populated
closer to the surface of the cloud in likely more diffuse and
spatially extended gas \citep[e.g.,][]{Abgrall1992}, it is reasonable
to expect that the $b$ values for higher rotational levels will be
comparable to or exceed those of the lower levels. Thus, to obtain
physical results for the column densities for $J = 2$ and $3$, we fit
all $J$ levels again imposing the prior that the probability is zero
for Doppler parameters less than $b_{J - 1} - \sigma_{b_{J - 1}}$ and
greater than $10$ km\,s$^{-1}$.

\begin{figure*}
  \centering
  \includegraphics[scale = 1.1]{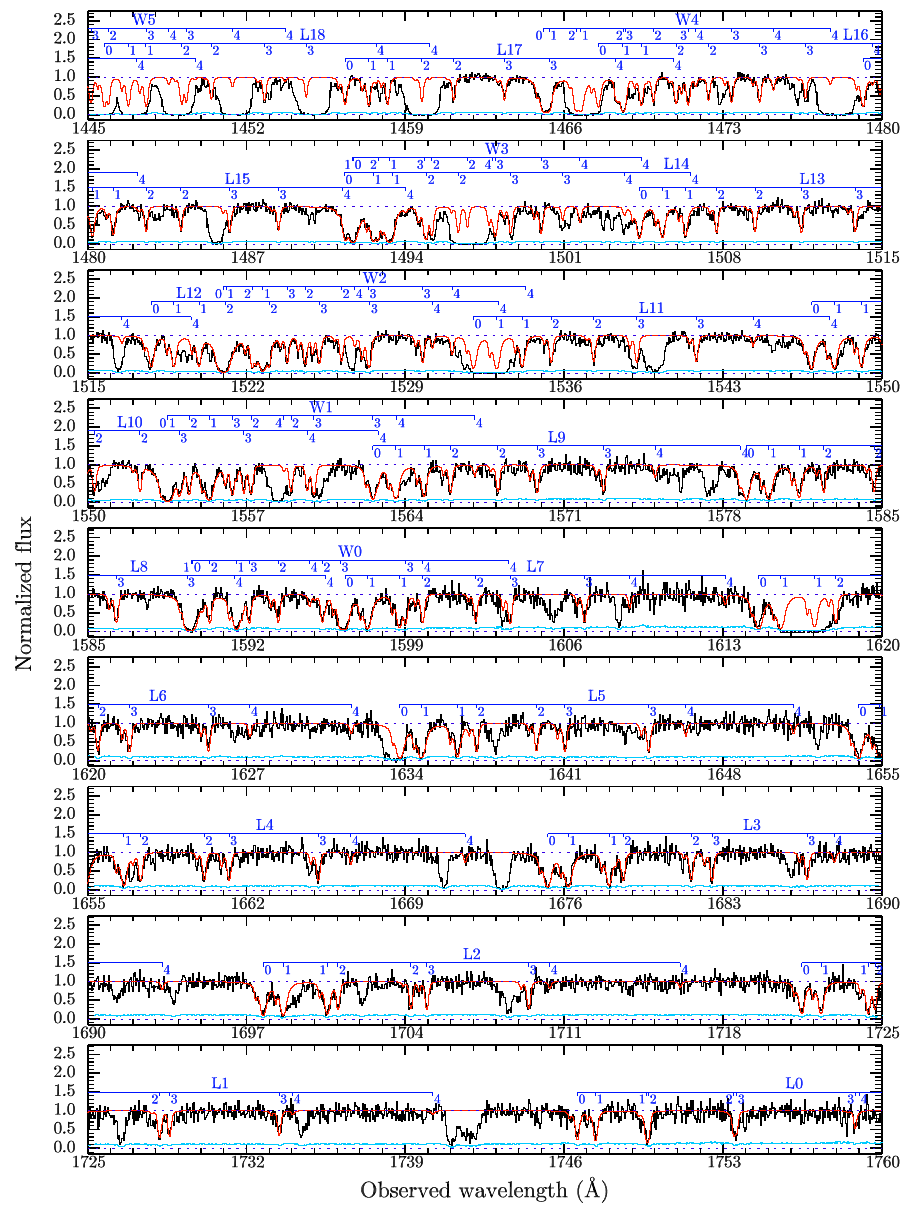}
  \caption{Model H$_{2}$ spectrum (red) overplotted on the observed COS
  spectrum towards J$0111\!-\!0316$ (black). The error spectrum is
  shown in cyan. The COS spectrum is rebinned to three pixels, or
  $\Delta\,\varv \approx 8$ km\,s$^{-1}$. Lyman (L) and Werner (W)
  band transitions for $J = 0 - 4$ are marked in blue; numbers below
  (above) the horizontal brackets indicate lower rotational (upper
  vibrational) quantum numbers of the H$_{2}$ transitions.}
  \label{fig:H2}
\end{figure*}

We give the Voigt profile fitting results in Table \ref{tab:H2}, and
we compare the best-fit model to the data in Fig.~\ref{fig:H2}; these
results yield a monotonic increase in $b$ value with $J$ level. If the
line width is dominated by thermal motions, the $b \approx 5$
km\,s$^{-1}$ observed at low $J$ level implies a temperature $T \sim 3
\times 10^{3}$ K. At high $J$ level ($b \approx 10$ km\,s$^{-1}$),
this becomes $T \sim 10^{4}$ K. Since this significantly exceeds the
temperature expected for diffuse molecular gas, the line width appears
to be dominated by turbulence in both cases, with turbulent motions a
factor of two higher at high rotational state compared to low. We
impose the same prior on the $b$ values for the H$_{2}$ absorption
observed in c4, but the large uncertainties associated with this weak
component make conclusions about possible trends of $b$ value with $J$
level challenging.

The total H$_{2}$ column density in the c6 (c4) component is
log[$N$(H$_{2}$)/cm$^{-2}$] $= 18.97\substack{+0.05 \\ -0.06}$
(log[$N$(H$_{2}$)/cm$^{-2}$] $= 15.8\substack{+0.5 \\ -0.3}$). Since
we cannot resolve the \ion{H}{1} absorption associated with these
velocity components, we calculate the H$_{2}$ fraction,
$f_{\text{H}_{2}} = 2N(\text{H}_{2})/[N(\text{\ion{H}{1}}) +
  2N(\text{H}_{2})] = 0.11\substack{+0.06 \\ -0.04}$, by attributing
all of the detected \ion{H}{1} to the H$_{2}$-bearing gas. This is
supported by the observation that $> 90\%$ of the \ion{O}{1} column
density is found in the central, H$_{2}$-bearing component (c6). This
H$_{2}$ fraction is among the highest $f_{\text{H}_{2}}$ observed for
systems of comparable $N$(\ion{H}{1}) in both the local and
high-redshift Universe \citep[see, e.g.,][and references
  therein]{Crighton2013, Muzahid2015}.

\begin{deluxetable}{lcccc}
  \label{tab:H2}
  \tablecaption{Molecular Hydrogen Properties}
  \tablehead{
    \colhead{} &
    \colhead{$\Delta\,\varv$}  &
    \colhead{} &
    \colhead{} &
    \colhead{$b$} \\
    \colhead{Comp.} &
    \colhead{(km\,s$^{-1}$)}  &
    \colhead{$J$} &
    \colhead{log[$N_{J}$/cm$^{-2}$]} &
    \colhead{(km\,s$^{-1}$)}
  }
  \startdata
  c4 & $-59.6$ & $0$ & $15.1\substack{+1.2 \\ -0.6}$ & $3\substack{+2 \\ -1}$ \\
   $ $ & & $1$ & $15.3\substack{+0.4 \\ -0.1}$ & $5\substack{+2 \\ -1}$\\
   $ $ & & $2$ & $14.4 \pm 0.2$ & $7 \pm 2$\\
   $ $ & & $3$ & $14.74 \pm 0.06$ & $8 \pm 2$\\
   $ $ & & $4$ & $< 13.9$ & $9$\tablenotemark{a} \\
   $ $ & & \textbf{Total} & $15.8\substack{+0.5 \\ -0.3}$ &  \\
  &   &   & \\
  c6 & $0$ & $0$ & $18.61\substack{+0.07 \\ -0.08}$ & $4.4\substack{+0.6 \\ -0.7}$\\
   & & $1$ & $18.71\substack{+0.08 \\ -0.09}$ & $5.3 \pm 0.3$ \\
   & & $2$ & $16.6\substack{+0.6 \\ -0.4}$ & $7 \pm 1$ \\
   & & $3$ & $16.1\substack{+0.4 \\ -0.2}$ & $8.0\substack{+0.9 \\ -1.3}$ \\
   & & $4$ & $14.64 \pm 0.04$ & $9.1\substack{+0.6 \\ -1.0}$ \\
   & & $5$ & $< 14.3$ & $10$\tablenotemark{a} \\
  $ $ & & \textbf{Total} & $18.97\substack{+0.05 \\ -0.06}$ \\
  \noalign{\smallskip}
  \enddata
  \tablenotetext{a}{$b$ values for upper limits are fixed based on the trend of $b$ with $J$ for lower $J$ levels.}
\end{deluxetable}

\subsubsection{H$_{2}$ excitation}
\label{sec:H2_ex}

\begin{figure*}
  \centering
  \includegraphics[scale = 0.5]{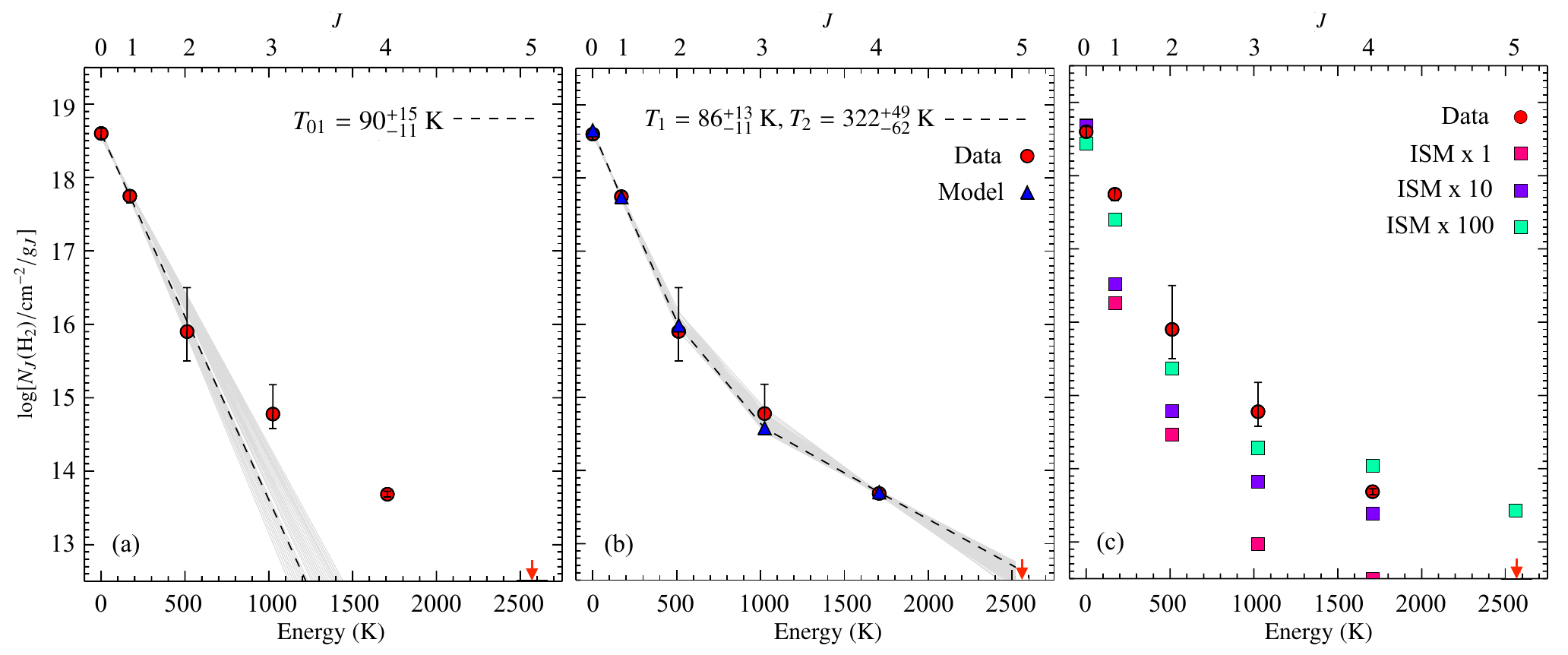}
  \caption{\textbf{(a)} H$_{2}$ column density as a function of lower
    rotational level, $J$, for the primary velocity component at
    $\Delta\,\varv = 0$ km\,s$^{-1}$. The temperature implied by the
    $J = 0$ and $J = 1$ column densities, $T_{01} = 90^{+15}_{-11}$ K,
    underpredicts the column density at $J \ge 3$. \textbf{(b)} A
    two-temperature model with $T_{1} = 86\substack{+13 \\ -11}$ K and
    $T_{2} = 322\substack{+49 \\ -62}$ K well reproduces the observed
    excitation state of the molecular hydrogen, but requires high gas
    density ($n(\text{H}) \gtrsim 10^{4}$ cm$^{-3}$) and thus very
    small cloud size ($l \sim 10^{-3}$ pc). In panels (a) and (b), the
    gray lines show 50 best-fit models where the $T_{01}$ and T$_{2}$
    values are drawn uniformly from the cumulative distribution
    functions for these parameters within the 68$\%$ confidence
    intervals, respectively. The dashed lines indicate the models with
    the maximum likelihood. \textbf{(c)} \texttt{CLOUDY} models of the
    excitation state of the H$_{2}$ in a radiative pumping scenario,
    where the ambient radiation field is equivalent to that observed
    locally, adjusted by a scale factor, $\alpha$. If the excitation
    state of the H$_{2}$ is primarily due to pumping, then the scale
    factor must be in the range $\alpha = 10 - 100$. The models shown
    here are for a two-sided geometry with constant gas pressure.}
  \label{fig:temp}
\end{figure*}

The H$_{2}$ column density distribution as a function of rotational
level gives insight into the excitation state of the molecular
hydrogen. We focus our excitation analysis on c6 due to the weak
nature and thus large associated uncertainties of c4. As shown in the
left panel of Fig.~\ref{fig:temp}, the H$_{2}$ column densities in the
$J = 0$ and 1 states of c6 imply an excitation temperature of
  $T_{01} = 90^{+15}_{-11}$ K assuming a Boltzmann
distribution. However, the H$_{2}$ column densities at $J > 2$ are not
well explained by this single temperature model; instead, these column
densities significantly exceed their predicted values, with increasing
discrepancy at higher $J$ level. The observed column densities can be
well reproduced by a two-temperature model with $T_{1} =
  86\substack{+13 \\ -11}$ K and $T_{2} = 322\substack{+49 \\ -62}$ K
(see the middle panel of Fig.~\ref{fig:temp}). The best-fit
temperatures and associated uncertainties for the one- and two-phase
models are determined by a maximum likelihood analysis; in the latter
case, the upper limit on $N_{J = 5}$ provides a strong constraint on
$T_{2}$. The failure of a single temperature model to reproduce the
excitation state of the molecular hydrogen is commonly noted in
absorbers at a range of redshifts \citep[e.g.,][]{Spitzer1975,
  Noterdaeme2007, Rawlins2018, Balashev2019}, and the observed
$T_{01}$ is consistent with the excitation temperatures seen in
H$_{2}$-bearing DLAs and sLLSs of comparable $N$(H$_{2}$) \citep[see,
  e.g.,][and references therein]{Muzahid2015}.

The presence of multi-phase gas produced by shocks or turbulent mixing
layers \citep[e.g.,][]{Gry2002, Ingalls2011} has been attributed to
the excitation of H$_{2}$-bearing gas observed in the Milky Way
\citep[e.g.,][]{Jenkins1997} and at higher redshift
\citep[e.g.,][]{Noterdaeme2007}. However, excitation of the $J > 2$
states via collisions in a thermalized medium requires relatively high
gas density ($n(\text{H}) \gtrsim 10^{4}$ cm$^{-3}$) and thus very
small absorber size ($l \sim 10^{-3}$ pc for the observed gas
column). Although there is evidence for sub-pc sizes among known
H$_{2}$-bearing clouds \citep[e.g.,][]{Reimers2003}, the very small
clump size required here implies an extremely low covering fraction of
these H$_{2}$ absorbers, making it unlikely to detect one in the 15
CUBS sightlines.

Alternative explanations for the elevated excitation state include
pumping by an ambient UV radiation field \citep{Jura1975} and direct
formation on high rotational levels \citep[e.g.,][]{Lacour2005}. Here
we explore the possibility that the excitation state of the H$_{2}$ is
due to UV pumping by estimating the required radiation field using the
spectral sythesis code \texttt{CLOUDY} (v13.05;
\citealt{Ferland2013}). Using the observed $N(\text{H}_{2})$ of the
stronger molecular absorber as the stopping condition, we run a series
of models at the total hydrogen volume density required to reproduce
the observed $N(\text{\ion{H}{1}})$ and $N(\text{H}_{2})$
simultaneously in the presence of the chosen radiation field.

We assume a constant gas pressure across the cloud and impose a
two-sided, plane-parallel geometry by modeling half of the observed
$N(\text{H}_{2})$ and $N(\text{\ion{H}{1}})$ and mirroring the
result. The models include grains consistent with the size
distribution and abundance of the local Galactic ISM, with the
abundance of both metals and grains reduced to 40$\%$ of the solar
neighborhood values (see \S\ref{sec:chem_abundance}). We apply the
Galactic cosmic ray background, which uses the H$^{0}$ and H$_{2}$
ionization rates of \citet{Indriolo2007} and \citet{Glassgold1974},
respectively. We use two ambient radiation fields available in
\texttt{CLOUDY} in addition to the cosmic microwave background: the
HM05 UV background at $z = 0.576$ (a 2005 update to the
Quasars+Galaxies field of \citealt{Haardt2001}) and the unextinguished
local ISM radiation field \citep{Black1987}, which is dominated by
starlight at the wavelengths relevant to H$_{2}$ excitation and
dissociation. We adjust the latter radiation field by a scale factor
ranging from $\alpha = 0\!-\!100$. The suite of models considered here
suggest that the \ion{H}{1} resides in a largely neutral zone, with
$> 90\%$ of the $N$(\ion{H}{1}) arising from regions with H ionization
fractions $< 1\%$.

We compare the observed and modeled $N_{J}$ as a function of $J$ level
for a range of scale factors, $\alpha$, in the right panel of
Fig.~\ref{fig:temp}. A scale factor of $\alpha = 10 - 100$ roughly
reproduces the excitation state of the H$_{2}$ at $J \ge 3$,
suggesting that a significantly elevated radiation field compared to
that estimated locally is required to explain the excitation state of
the molecular hydrogen if radiative pumping is the dominant
mechanism. For $\alpha = 10$, the model produces an H$_{2}$ volume
density of $n(\text{H}_{2}) \approx 3 \times 10^{2}$ cm$^{-3}$ (as
compared to a total $n(\text{H}) \approx 2 \times 10^{3}$ cm$^{-3}$)
and a kinetic temperature of $T_{\text{kin}} = 35$ K in the cloud
core. At $\alpha = 100$, this becomes $n(\text{H}_{2}) \approx 7
\times 10^{3}$ cm$^{-3}$ ($n(\text{H}) \approx 3 \times 10^{4}$
cm$^{-3}$) and $T_{\text{kin}} = 70$ K. In the core of the cloud,
these kinetic temperatures are comparable to the modeled $T_{01}$,
consistent with significant self-shielding. The difference between the
observed and modeled $T_{01}$ may be due to a range of uncertainties
in the model, including the shape of the incident radiation field and
the relative geometries of the absorber and any local source(s) of UV
photons.

We also produce alternative models imposing constant total hydrogen
density, as well as both one- and two-sided geometries for the
constant density and constant gas pressure cases. These models lead to
the same general conclusion that a radiative pumping scenario requires
an ambient radiation field elevated above that observed in the solar
neighborhood by one to two orders of magnitude. Allowing for
$f_{\text{H}_{2}} > 11\%$, as may be the case if not all of the
observed $N(\text{\ion{H}{1}})$ is associated with the molecular
absorber, does not significantly alter these conclusions. In this
picture, the rise in $b$ value with $J$ level follows naturally from a
stratified cloud in which the $J = 0$ and $1$ levels are primarly
populated closer to the interior of the cloud than the higher
rotational levels and thus may arise from less turbulent and spatially
extended gas \citep[e.g.,][]{Abgrall1992, Balashev2009}.

\subsection{Chemical abundance pattern}
\label{sec:chem_abundance}

We quantify the metallicity of the H$_{2}$-bearing DLA using the
available, uncontaminated \ion{O}{1} transitions in the COS data,
including \ion{O}{1}\,$\lambda$922, 936, and 1039, where the weak
$\lambda$922 line is unsaturated for all components. As shown in
Fig.~\ref{fig:HI_H2_metals} and Table \ref{tab:metals}, there is
detectable \ion{O}{1} absorption associated with velocity components
c4, c6, and c9. The \ion{O}{1} column density,
log[$N$(\ion{O}{1})/cm$^{-2}$] $= 16.5$, yields a metallicity of
$\approx 50$\% of solar. Since the \ion{H}{1} column density
associated with individual metal-line components is unknown, this is a
velocity-integrated metallicity that represents an average value for
the absorber.

\begin{figure*}
  \centering
  \includegraphics[scale = 0.5]{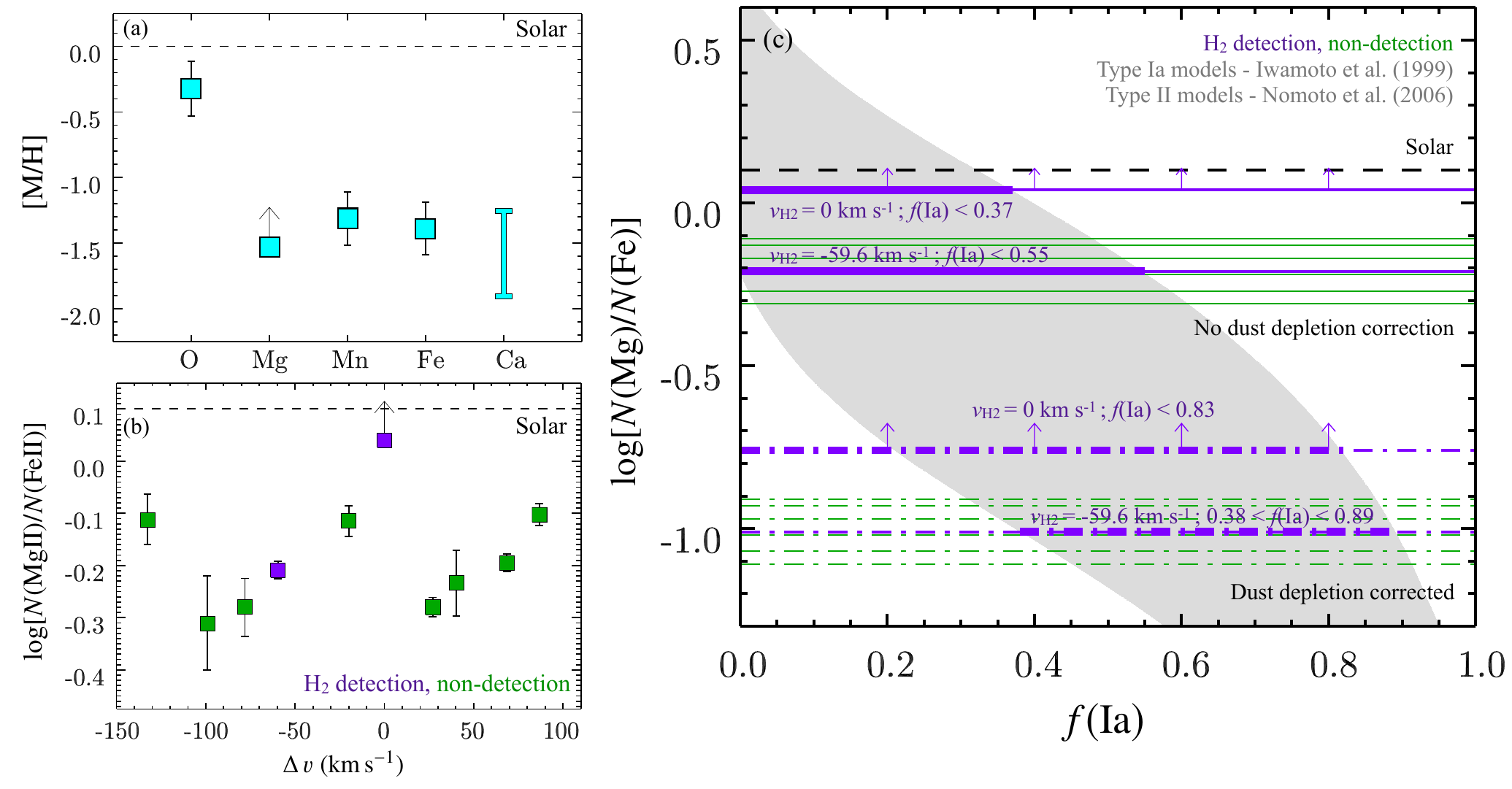}
  \caption{\textbf{(a)} Velocity-integrated metallicities, [M/H], for
    the increasingly refractory elements O, Mg, Mn, Fe, and Ca. The
    larger departures from solar metallicity among the more refractory
    elements suggest the presence of dust depletion. The range of
    values shown for [Ca/H] follows from the range of ionization
    corrections implied by the suite of \texttt{CLOUDY} models
    explored in \S\ref{sec:H2_ex}. \textbf{(b)} Observed
    log[$N$(\ion{Mg}{2})/$N$(\ion{Fe}{2})] as a function of
    $\Delta\,\varv$ for all velocity components. Broad consistency in
    the column density ratios may suggest a shared history for some
    subsets of components, but variation in chemical abundance and/or
    dust depletion pattern is evident across the system. \textbf{(c)}
    Observed log[$N$(Mg)/$N$(Fe)] (horizontal lines) compared to the
    range of model predictions (gray shading) for a given Type Ia SNe
    fraction, $f(\text{Ia})$, derived from the yield models of
    \citet{Iwamoto1999} (Type Ia) and \citet{Nomoto2006} (Type
    II). Each horizontal line indicates a different velocity
    component, with H$_{2}$-bearing (-free) absorbers indicated in
    purple (green). The observed abundance ratios, shown as solid
    lines, are consistent with moderate to no contribution from Type
    Ia SNe. In dashed lines, we correct the observed ratios for the
    dust depletion pattern of \citet{DeCia2016} for a depletion factor
    of [Zn/Fe] = 1.2, chosen to approximate a warm galactic disk
    \citep{Savage1996}. Once depletion corrected, the H$_{2}$-bearing
    component associated with c4 ($\Delta\,\varv = -59.6$
    km\,s$^{-1}$) is consistent with significant contributions from
    Type Ia SNe ($0.38 < f(\text{Ia}) < 0.89$).}
  \label{fig:chem_abundance}
\end{figure*}

Interesting trends emerge when the chemical abundance pattern of other
atomic species is considered. In panel (a) of
Fig.~\ref{fig:chem_abundance}, we plot the velocity-integrated
metallicity, [M/H], of five atomic species: O (traced by
$N$(\ion{O}{1})), Mg ($N$(\ion{Mg}{1}) + $N$(\ion{Mg}{2})), Mn
($N$(\ion{Mn}{2})), Fe ($N$(\ion{Fe}{2})), and Ca
($N$(\ion{Ca}{2})). The suite of \texttt{CLOUDY} models produced in
\S\ref{sec:H2_ex} suggest that the \ion{H}{1} arises in a neutral zone
with H ionization fraction $< 1\%$, and thus we expect \ion{O}{1},
\ion{Mg}{2}, \ion{Mn}{2}, and \ion{Fe}{2} to be the dominant
ionization state for each respective element. We use the
\texttt{CLOUDY} models to corroborate that the contributions from
other ionization states are negligible. Given the relatively low
second ionization potential of Ca (11.9 eV), an ionization correction
is required. We find $N({\text{\ion{Ca}{2}}})/N(\text{Ca}) = 0.2 -
0.9$ for the range of assumptions explored in our
photoionization models (one- vs. two-sided, constant gas pressure
vs. density, $\alpha = 0 - 100$), and we thus display [Ca/H] as a band
determined from this range of ionization corrections.

In the upper left panel of Fig.~\ref{fig:chem_abundance}, the atomic
species are ordered from least to most refractory. O, the least
refractory element considered here, is expected to experience minimal
depletion onto dust grains in the Milky Way \citep{Savage1996} and
higher redshift DLA environments \citep[e.g.,][]{DeCia2016}. In
contrast, in a dusty medium, we expect increasingly significant
depletion of Mg, Mn, Fe, and Ca, and indeed we observe larger
departures from solar metallicity for more refractory elements. We
estimate the dust-to-gas ratio, $\kappa_{\text{X}} =
10^{\text{[X/H]}}(1 - 10^{\text{[Fe/X]}})$, by taking X = O as the
element least likely to be affected by dust depletion. This yields a
velocity-integrated dust-to-gas ratio of $\kappa_{\text{O}} \approx
0.4$, and it is possible that this ratio is even higher in the
velocity components that are H$_{2}$-bearing. This implies the
presence of dust in the absorber, consistent with findings that higher
H$_{2}$ fractions tend to be found in systems with higher dust-to-gas
ratios, although additional effects including the gas density and
ambient radiation field are also relevant \citep[e.g.,][]{Ledoux2003}.

We can ask to what extent variation in the chemical abundance pattern
is observed between different velocity components. In the absence of
kinematically-resolved \ion{H}{1} measurements, we examine the ratio
of the \ion{Mg}{2} and \ion{Fe}{2} column densities,
log[$N$(\ion{Mg}{2})/$N$(\ion{Fe}{2})], as a function of velocity,
$\Delta\,\varv$, in panel (b) of
Fig.~\ref{fig:chem_abundance}. Firstly, we note that all velocity
components are consistent with sub-solar
log[$N$(\ion{Mg}{2})/$N$(\ion{Fe}{2})], falling below the solar value
by $\approx 0.05 - 0.4$ dex. Secondly, component-to-component
variation in this column density ratio suggests a non-uniform chemical
abundance and/or dust depletion pattern across the system.

However, some subsets of components have consistent
log[$N$(\ion{Mg}{2})/$N$(\ion{Fe}{2})] within the errors. Notably, the
H$_{2}$-bearing component associated with c4 is consistent with c2,
c3, c8, and c9 within $\lesssim 1\sigma$. Relative homogeneity in the
component-to-component chemical properties and ionization states of
absorbers has been interpreted in other systems as arising from clouds
sharing similar physical conditions and environments
\citep[e.g.,][]{Lopez2002, Prochaska2003}. For the H$_{2}$-bearing
component associated with c6, we find
log[$N$(\ion{Mg}{2})/$N$(\ion{Fe}{2})] $> 0.04$; as this exceeds the
ratio observed for all other velocity components, we cannot infer that
the stronger molecular absorber originates from the same astrophysical
processes as the other velocity components based on their respective
column density ratios.

In panel (c) of Fig.~\ref{fig:chem_abundance}, we again show the
observed log[$N$(Mg)/$N$(Fe)] for each velocity component compared to
solar. These results suggest moderate Fe enhancement with respect to
the solar abundance pattern, as is often associated with preferential
enrichment from Type Ia supernovae (SNe). In the same panel, we
indicate the range of log[$N$(Mg)/$N$(Fe)] expected for a given Type
Ia fraction, $f(\text{Ia}) = N_{\text{Ia}}/(N_{\text{Ia}} +
N_{\text{II}})$, where $N_{\text{Ia}}$ and $N_{\text{II}}$ are the
number of Type Ia and Type II events that have contributed to
chemically enriching the gas, respectively. The range of model
predictions for a given $f(\text{Ia})$ is due to the spread in
chemical abundance pattern corresponding to a range of initial
metallicities, deflagration speeds, and densities at ignition (Type
Ia) and progenitor masses (Type II; $M_{\text{proj}} = 13 -
40~\text{M}_{\odot}$). The Type Ia yields are taken from
\citet{Iwamoto1999}, and the Type II yields are adopted from
\citet{Nomoto2006} for $Z = 0.004$.

As shown by solid lines in panel (c) of Fig.~\ref{fig:chem_abundance},
the observed values of log[$N$(Mg)/$N$(Fe)] for the two
H$_{2}$-bearing absorbers suggest $f(\text{Ia}) < 0.37$ (c6) and
$f(\text{Ia})$ < 0.55 (c4). Thus, taken at face value, the observed
log[$N$(Mg)/$N$(Fe)] is consistent with no enrichment by Type Ia
SNe. However, as discussed above, the growing departure of [M/H] from
solar metallicity for increasingly refractory elements strongly
suggests the presence of dust depletion. As shown by dashed lines in
the same Figure, we overplot log[$N$(Mg)/$N$(Fe)] corrected for the
depletion pattern of \citet{DeCia2016} for a depletion factor of
[Zn/Fe] = 1.2, chosen to approximate the warm disk of the Milky Way
\citep{Savage1996}. For the H$_{2}$-bearing component associated with
c4, this suggests $0.38 < f(\text{Ia}) < 0.89$, implying significant
contributions from Type Ia SNe. Raising the metallicity of the Type II
yield models of \citet{Nomoto2006} to $Z = 0.02$ has little effect on
the results; a choice of $Z = 0.001$ increases the upper bound on the
suggested $f(\text{Ia})$ by $\lesssim 50\%$.

Given our lack of knowledge of the intrinsic chemical abundance and
depletion patterns, we cannot definitively determine the presence and
extent of Fe enhancement in the absorbing gas. The extent of the
depletion is highly uncertain; depletion factors ranging from [Zn/Fe]
$\sim 0.5 - 1.5$ are observed in DLAs with metallicities near solar
\citep{Zou2018}. However, it is clear that simultaneously explaining
all chemical abundance ratios strongly suggests the influence of both
dust depletion and Fe enhancement, the latter of which is seen in
systems where Type Ia SNe have made significant contributions to the
chemical enrichment history.

\subsection{The galaxy environment of the DLA}
\label{sec:group}

\begin{figure*}
  \centering
  \includegraphics[scale = 1.3]{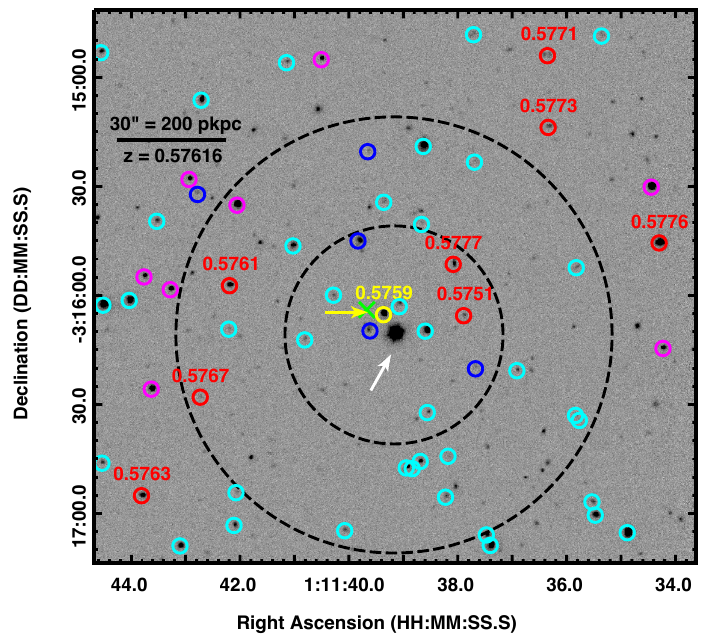}
  \caption{Magellan $H$-band image of the field of J$0111\!-\!0316$,
  indicated by the white arrow. Galaxies with secure, multi-line
  redshifts within $d = 600$ pkpc and $\Delta\,z = \pm 0.005$ of the
  DLA at $z = 0.576$ are marked by red circles. Galaxies with multi-
  and single-line redshifts outside of this velocity range are indicated in
  cyan and blue, respectively. Stars in the field are marked by
  magenta circles. The massive, early-type galaxy at the closest
  impact parameter, $d = 41$ pkpc, is shown in yellow, and the center
  of mass of this galaxy and the four additional galaxies projected
  within its virial radius is indicated by the green cross.  The
  dashed black circles have radii of $30''$ and $60''$, or $\approx
  200$ pkpc and $400$ pkpc at the redshift of the absorber.}
  \label{fig:galaxies}
\end{figure*}

\begin{figure}
\includegraphics[scale = 0.45]{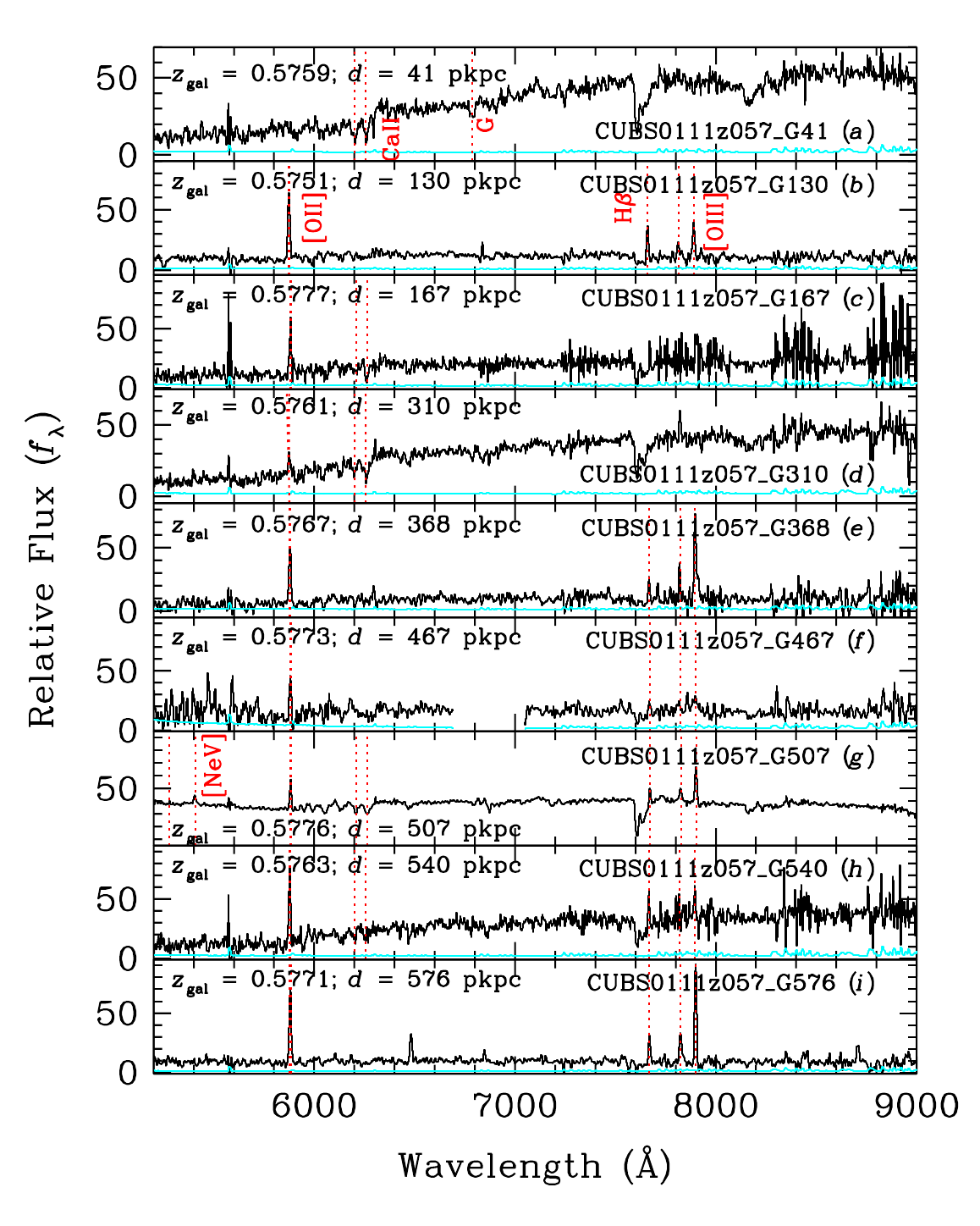}
\caption{Magellan spectroscopy of the nine galaxies within $d = 600$ pkpc and $\Delta\,z = \pm 0.005$ of the DLA at $z = 0.576$. The two galaxies that are closest to the QSO in projected distance have contrasting spectra; in panel \textbf{(a)}, the closest galaxy at $d = 41$ pkpc is a massive, absorption-line dominated system, while in panel \textbf{(b)}, the galaxy at $d = 130$ pkpc has a star-forming, emission-line dominated spectrum.}
\label{fig:gal_spec}
\end{figure}

The Magellan galaxy redshift survey reveals nine galaxies with secure,
multi-line redshifts within an impact parameter of $d = 600$ pkpc and
$\Delta\,\varv_{g}$ = 300 km\,s$^{-1}$ of the H$_{2}$-bearing
absorber. We show the locations of these galaxies on a Magellan
$H$-band image of the field and their Magellan spectroscopy in
Figs.~\ref{fig:galaxies} and \ref{fig:gal_spec}, respectively. In
Table \ref{tab:galaxies}, we tabulate their RA and decl. offsets
($\Delta\,\alpha$, $\Delta\,\delta$), angular separations ($\theta$),
and impact parameters ($d$) from the QSO. We also list their redshifts
($z$), velocity offsets ($\Delta\,\varv_{\text{g}}$), apparent and
absolute $r$-band magnitudes ($m_{r}$, $M_{r}$), stellar masses
($M_{\text{star}}$), and spectral types. Beyond the nine galaxies
identified here, the next closest galaxy in projection is at $d
\approx 1650$ pkpc within $\Delta\,z = \pm 0.005$ of the absorber
redshift. The uncertainty on the galaxy redshift measurements is
$\Delta z \approx 0.0002$ ($\Delta z \approx 0.0003$) for systems
dominated by emission (absorption).

The galaxy at the closest projected distance to the QSO, $d = 41$
pkpc, is also the most massive, with
log$(M_{\text{star}}/\text{M}_{\odot}) = 10.9$. We estimate the
stellar mass from the galaxy photometry following
\citet{Johnson2015}. As shown in panel (a) of Fig.~\ref{fig:gal_spec},
CUBS0111z057\_G41 (hereafter G41) has an absorption-line dominated
spectrum, suggesting that it is a massive, early-type system. We place
a 2$\sigma$ upper limit on its star formation rate of $< 0.2$
$\text{M}_{\odot}$ yr$^{-1}$ based on the absence of
[\ion{O}{2}]\,$\lambda\lambda$3727, 3729 emission in its spectrum
\citep{Kewley2004}.  Using the fitting code
\texttt{bagpipes}\footnote{https://bagpipes.readthedocs.io/en/latest/}
\citep{Carnall2018}, we find the age of the stellar population to be
$\approx 3.2$ Gyr assuming a single burst of star formation. A tau
model (SFR $\propto \exp(-t/\tau)$) produces a similar age of $\approx
3.4$ Gyr. This galaxy is offset in velocity from c6 by
$\Delta\,\varv_{\text{g}} = -57$ km\,s$^{-1}$.

At over three times the projected distance, the next closest galaxy
($d = 130$ pkpc; CUBS0111z057\_G130, hereafter G130) has a
contrasting, emission-line dominated spectrum (see panel (b) of
Fig.~\ref{fig:gal_spec}). This star-forming galaxy has an unobscured
star formation rate of $\sim 1$ $\text{M}_{\odot}$ yr$^{-1}$ inferred
from its [\ion{O}{2}] equivalent width \citep{Kewley2004}. Another
galaxy of note, CUBS0111z057\_G507 (hereafter G507), is the brightest
in the field and is located at $d = 507$ pkpc. As seen in panel (g) of
Fig.~\ref{fig:gal_spec}, the clear detection of
[\ion{Ne}{5}]\,$\lambda$3426 emission and a power-law term
superimposed on the stellar continuum indicate the presence of an
AGN. From Equation 12 of \citet{Reyes2008}, we find a UV magnitude of
$M$(2500) = -22.1 at 2500 \AA\ from the observed
[\ion{O}{3}]\,$\lambda$5008 luminosity of the AGN. This suggests that
any radiation field due to the AGN at the location of the DLA is
likely sub-dominant compared to the UV background at the wavelengths
relevant to the H$_{2}$ excitaton. G130 and G507 are offset in
velocity from c6 by $\Delta\,\varv_{\text{g}} = -209$ km\,s$^{-1}$ and
$+266$ km\,s$^{-1}$, respectively.

It is natural to ask whether these nine galaxies clustered within $d =
600$ pkpc and with velocity dispersion $\sigma_{\text{gal}} = 152$
km\,s$^{-1}$ are members of a bound group. Beginning with the most
massive galaxy, G41, we use the estimated stellar mass,
log$(M_{\text{star}}/\text{M}_{\odot}) = 10.9$, to infer the total
dynamical mass, log$(M_{\text{halo}}/\text{M}_{\odot}) = 12.9$, from
the stellar mass - halo mass relation of \citet{Behroozi2019} at the
redshift of the DLA. We then calculate the corresponding virial
radius, $r_{\text{vir}} \approx 370$ kpc, using the formalism of
\citet{Maller2004}, who define $r_{\text{vir}}$ based on the
redshift-dependent overdensity criterion of \citet{Bryan1998}. This
virial radius encloses four additional galaxies in projection, G130,
CUBS0111z057\_G167, CUBS0111z057\_G310, and
CUBS0111z057\_G368. Repeating this process with the summed masses of
all five galaxies yields a comparable $r_{\text{vir}}$ that does not
encompass any additional galaxies. A similar analysis for the next two
most massive galaxies, G507 and CUBS0111z057\_G310, finds no
additional galaxies within the virial radii of these systems. This
suggests that the nine galaxies do not represent a single bound group,
but instead indicate a loose association in which G41 may be bound to
its four closest projected neighbors. We discuss the possible
association of the damped absorber with these galaxies in the
following Section.

\begin{longrotatetable}
\begin{deluxetable*}{llrrrrccccrcc}
  \tablecaption{Properties of galaxies in the vicinity of the DLA at $z = 0.576$}
  \tablehead{
    \colhead{} &
    \colhead{} &
    \colhead{$\Delta\,\alpha$\tablenotemark{a}} &
    \colhead{$\Delta\,\delta$} &
    \colhead{$\theta$} &
    \colhead{$d$} &
    \colhead{} &
    \colhead{$\Delta\,\varv_{\text{g}}$\tablenotemark{b}} &
    \colhead{$m_{r}$\tablenotemark{c}} &
    \colhead{$M_{r}$\tablenotemark{d}} &
    \colhead{} &
    \colhead{} \\
    \colhead{ID} &
    \colhead{Name} &
    \colhead{($''$)} &
    \colhead{($''$)} &
    \colhead{($''$)} &
    \colhead{(pkpc)} &
    \colhead{$z$} &
    \colhead{(km\,s$^{-1}$)} &
    \colhead{(mag)} &
    \colhead{(mag)} &
    \colhead{log$(M_{\text{star}}/\text{M}_{\odot})$\tablenotemark{e}} &
    \colhead{Type\tablenotemark{f}}
  }
  \startdata
  J$011139.36-031605.3$ & CUBS0111z057\_G41 & $+2.9$ & $+5.5$ & 6.3 & 41.0 & 0.5759 & $-57$ & $21.47\pm0.03$ & $-22.2$ & 10.9 & Abs \\
  J$011137.89-031605.6$ & CUBS0111z057\_G130 & $-19.1$ & $+5.3$ & 19.9 & 130.1 & 0.5751 & $-209$ & $22.52\pm0.12$ & $-20.1$ & 9.4 & Em\\
  J$011138.08-031551.4$ & CUBS0111z057\_G167 & $-16.3$ & $+19.4$ & 25.4 & 166.5 & 0.5777 & $+285$ & $22.43\pm0.09$ & $-20.5$ & 10.0 & Abs/Em \\
  J$011142.19-031557.3$ &  CUBS0111z057\_G310 & $+45.3$ & $+13.6$ & 47.3 & 310.1 & 0.5761 & $-19$ & $21.90\pm0.03$ & $-21.6$ & 10.4 & Abs/Em \\
  J$011142.73-031628.0$ & CUBS0111z057\_G368 & $+53.3$ & $-17.1$ & 56.0 & 367.6 & 0.5767 & $+95$ & $23.95\pm0.06$ & $-18.7$ & 8.9 & Em \\
  J$011136.34-031513.8$ & CUBS0111z057\_G467 & $-42.4$ & $+57.1$ & 71.1 & 466.9 & 0.5773 & $+209$ & $22.42\pm0.06$ & $-20.5$ & 9.8 & Em \\
  J$011134.30-031545.6$ & CUBS0111z057\_G507 & $-72.9$ & $+25.3$ & 77.2 & 506.9 & 0.5776 & $+266$ & $20.28\pm0.01$ & $-22.8$ & 10.7 & Abs/Em \\
  J$011143.81-031655.1$ & CUBS0111z057\_G540 & $+69.5$ & $-44.2$ & 82.3 & 540.0 & 0.5763 & $+19$ & $22.35\pm0.06$ & $-20.8$ & 10.2 & Abs/Em \\
  J$011136.35-031454.0$ & CUBS0111z057\_G576 & $-42.2$ & $+76.8$ & 87.7 & 575.5 & 0.5771 & $+114$ & $22.29\pm0.07$ & $-20.6$ & 9.8 & Em \\
  \enddata
  \tablenotetext{a}{RA offsets ($\Delta\,\alpha$), decl. offsets ($\Delta\,\delta$), angular separations ($\theta$), and impact parameters ($d$) are measured with respect to the QSO position at RA, decl. $=$ 01h11m39.170s, -03d16m10.89s.}\tablenotetext{b}{Velocity offsets ($\Delta\,\varv_{\text{g}}$) are measured with respect to the strongest metal-line absorption at $z = 0.57616$.}\tablenotetext{c}{$r$-band apparent magnitudes are from the Dark Energy Survey \citep{Abbott2018}, with the exception of the faintest galaxy, CUBS0111z057\_G368; for this system, $m_{r}$ is from Magellan photometry.}\tablenotetext{d}{The characteristic rest-frame absolute $r$-band magnitude is $M_{r,\text{star}} = -21.7$ ($M_{r,\text{star}} = -21.96$) for blue (red) galaxies following \citet{Cool2012}.}\tablenotetext{e}{We determine $M_{\text{star}}$ using the photometric fitting formulae of \citet{Johnson2015}. Note that for CUBS0111z057\_G507, whose spectrum suggests the presence of an AGN, we down-correct the estimated stellar mass by a factor of two, as $\gtrsim 50$\% of the $r$-band light appears to come from the compact nucleus.}\tablenotetext{f}{Spectral type: absorption-dominated (abs), emission-dominated (em), or composite (abs/em).}
  \label{tab:galaxies}
\end{deluxetable*}
\end{longrotatetable}

\section{Discussion}
\label{sec:diss}

We report the detection of an H$_{2}$-bearing DLA at $z = 0.576$ in a
well-characterized galaxy overdensity. The observed $f_{\text{H}_{2}}
= 0.11\substack{+0.06 \\ -0.04}$ is high among known systems of
comparable log[$N$(\ion{H}{1})/cm$^{-2}$] $\approx 20.1$ \citep[see,
  e.g.,][and references therein]{Crighton2013, Muzahid2015}, possibly
due to the relatively high dust-to-gas ratio ($\kappa_{\text{O}}
\approx 0.4$). This may also be due to ram pressure stripping or
photoevaporation of the \ion{H}{1} gas, as possibly observed at low
redshift in star-forming regions that are spatially offset from the
peak of the \ion{H}{1} gas distribution in interacting galaxies
\citep[e.g.,][]{MendesdeOliveira2004, Werk2008}. The absence of
associated \ion{O}{6} absorption, log[$N$(\ion{O}{6})/cm$^{-2}$] $<
14.0$, is notable. If \ion{O}{6} arises primarily at interfaces
between cool clouds and a hot, volume-filling phase, the lack of
\ion{O}{6} absorption suggests partial covering of such structures in
the relevant galaxy halo(s) and intra-group medium.

For the stronger of the two H$_{2}$-bearing velocity components, the
excitation state of the molecules may be due to UV pumping by an
ambient radiation field that is elevated above that observed in the
solar neighborhood by a factor of $\alpha = 10\!-\!100$. This field
may be produced by shocks arising from collisions between cool clouds
or at the interface between these clouds and the hot, ambient medium,
or may imply the presence of undetected star formation obscured by
dust or the glare of the QSO (see point III below). In this picture,
the observed rise in $b$ value with $J$ level can be attributed to
stratification in the cloud, in which $J > 1$ levels are primarily
populated closer to the surface of the cloud and thus in more diffuse
and spatially extended gas than the lower rotational levels
\citep[e.g.,][]{Abgrall1992, Balashev2009}.

The galaxy redshift survey presents an unprecedented opportunity to
explore the origin and physical conditions of an H$_{2}$-bearing DLA
in the context of its galactic environment. Nine galaxies are detected
within $d = 600$ pkpc and $\Delta\,\varv_{g}$ = 300 km\,s$^{-1}$ of
the absorber, and here we discuss the possible association of the
diffuse molecular gas and the relevant galaxies.

\textit{I. Association with a massive, early-type galaxy at $d = 41$
  pkpc.} Several lines of evidence favor an association between the
H$_{2}$-bearing DLA and the massive elliptical galaxy, G41. In
addition to having the smallest projected distance ($d = 41$ pkpc) and
a modest velocity offset ($\Delta\,\varv_{\text{g}} = -57$
km\,s$^{-1}$), this galaxy is also the most massive system observed at
the redshift of the DLA and represents almost $70\%$ of the stellar
mass of the small group comprised of this galaxy and its four closest
neighbors. Furthermore, as discussed in \S~\ref{sec:chem_abundance},
the chemical abundance pattern suggests Fe enhancement consistent with
significant contributions from Type Ia SNe. This chemical signature is
commonly seen in the inner halos of massive quiescent galaxies ($d
\lesssim 100$ pkpc), in contrast to their outer halos and star-forming
systems \citep[e.g.,][]{Zahedy2017}. The observed sub-solar
metallicity at $d \approx 40$ pkpc is not in tension with the super-solar
stellar metallicities often observed at the centers of massive
quiescent galaxies, due to the stellar metallicity gradients
observed in these systems \citep[e.g.,][]{Spolaor2010} and the
significant scatter seen within single luminous red galaxy halos
suggesting a range of gas cloud origins and poor gas mixing
\citep{Zahedy2019}. Given the observed projected distance, $d = 41$
pkpc, the absorber could either be associated with the outer ISM or
true CGM of this massive galaxy.

Significant amounts of neutral atomic and molecular gas have been
observed in early-type galaxies at low and intermediate redshifts. In
the local Universe, the ATLAS$^{\text{3D}}$ project yielded a $40\%$
detection rate of \ion{H}{1} 21-cm emission in non-cluster
environments in a volume-limited survey of early-type galaxies
\citep{Serra2012}; a $22\%$ detection rate of CO emission in both
cluster and non-cluster galaxies was found in a similar sample
\citep{Young2011}. On larger scales, extensive reservoirs of cool gas
have been detected in the CGM of intermediate-redshift, early-type
galaxies using QSO absorption-line probes
\citep[e.g.,][]{Gauthier2009, Bowen2011, Huang2016, Chen2018}. These
studies have demonstrated that luminous red galaxy halos can host
comparable amounts of cool gas as their star-forming counterparts
\citep[e.g.,][]{Zahedy2019}.

To our knowledge, direct detections of molecules at distances of
$\gtrsim 40$ pkpc from early-type galaxies in non-cluster environments
have not been reported in the literature. Indeed, published detections
of molecules on scales of many tens of pkpc have been confined to
interacting, star-forming galaxies in the local Universe
\citep[e.g.,][]{Appleton2006} and to proto-cluster environments at
high redshift \citep[e.g.,][]{Emonts2016, Emonts2019}. However, a lack
of detections on comparable scales around normal, early-type galaxies
at low and intermediate redshift is not surprising. In the local
Universe, single-dish CO surveys such as ATLAS$^{\text{3D}}$
\citep[e.g.,][]{Young2011} generally obtain single pointings, centered
on the galaxies, with beam sizes of a few pkpc FWHM; thus, they are
not designed to probe galaxy halos at large angular extents. At
intermediate redshift, state-of-the-art sub-mm interferometers such as
the Atacama Large Millimeter/submillimeter Array (ALMA) have detected
the dense molecular gas reservoirs associated with the central regions
of DLA host galaxies \citep[e.g.,][]{Kanekar2018}, but lack the
sensitivity to probe the diffuse molecular gas that may be found in
their outskirts.

However, there is indirect evidence that supports several scenarios
for the presence of H$_{2}$ at $d \approx 40$ pkpc from a massive,
early-type galaxy. In the local Universe, galaxy interactions appear
to enhance the incidence and physical extent of cold gas outside of
the densest group and cluster environments
\citep{Davis2011}. \citet{Serra2012} report a diverse range of \ion{H}{1}
morphologies in early-type galaxies, including evidence for tidal
tails and scattered clouds at tens of pkpc from the host
galaxy. Similar phenomena are seen in emission from cool ionized gas
\citep[e.g.,][]{Johnson2018}. Furthermore, such irregular \ion{H}{1}
morphologies are more commonly seen in denser environments, suggesting
a causal relationship with close companions. It is possible that
interactions between the early-type galaxy and member(s) of its
(small) group environment resulted in transport and/or compression of
gas in its hot halo, providing conditions for H$_{2}$ survival or in
situ formation \citep[e.g.,][]{Walter2006}.

Indeed, the Magellanic Stream is a local example of an extended
\ion{H}{1} gas distribution embedded in a hot halo that provides
conditions suitable for H$_{2}$ to survive and possibly to form
\citep[e.g.,][]{Richter2018}. Stephan's Quintet is another instance in
which a shock front produced by a galaxy traveling supersonically
through the hot intragroup gas can compress the ambient medium,
leading to molecule formation in the dense, rapidly-cooling phase
\citep{Appleton2006, Appleton2017}. In these scenarios, the apparent
Fe enhancement of the absorber suggests that the gas originates
primarily in the early-type galaxy or in a companion in which star
formation has not been active for a significant time. Additionally,
AGN activity can produce molecules at distances of tens of kpc from
host galaxies \citep[e.g.,][]{Cicone2014, Salome2016, Rudie2017},
either through direct transport, compression of the ambient medium, or
triggering its formation via cooling instabilities in outflows
\citep{Richings2018a, Richings2018b}. As there is no evidence for
current AGN activity in the massive, early-type galaxy, this model
requires the molecules to have survived from a past epoch of AGN
activity.

\textit{II. Association with a star-forming galaxy at $d = 130$ pkpc.}
The moderately star-forming galaxy (SFR $\sim 1$ $\text{M}_{\odot}$
yr$^{-1}$) at $d = 130$ pkpc is the next natural origin to consider
for the DLA. Star-formation driven outflows can transport molecules
into the CGM of late-type galaxies \citep[e.g.,][]{Rupke2019},
although these outflows typically occur on scales of kpc to tens of
kpc, in contrast to the $\sim 100$ kpc scale implied here
\citep[see][and references therein]{Cicone2014, Walter2017}. In
addition to the large physical scale, the chemical abundance pattern
of the H$_{2}$-bearing DLA presents a challenge to this picture. Fe
enhancement is not characteristic of the ISM or CGM of late-type
galaxies \citep[e.g.,][]{Zahedy2017}, and thus is not expected in
star-formation driven outflows.

\textit{III. Association with a galaxy hidden by the QSO.} It is
possible that the H$_{2}$-bearing gas arises in a galaxy at very small
impact parameter that is hidden by the bright background QSO ($d
\lesssim 4$ pkpc derived from the $\approx 0.6''$ seeing disk). At the
redshift of the DLA, we place a 3$\sigma$ upper limit on the
corresponding unobscured star formation rate of $\text{SFR} < 0.1$
$\text{M}_{\odot}$ yr$^{-1}$ based on an upper limit on the
[\ion{O}{3}]\,$\lambda$5008 equivalent width within a circular
aperature of radius $R = 0.6''$ following the relation of
\citet{Moustakas2006} for dwarf galaxies with $M_{\text{B}} > -16$. A
local source of UV photons from star-forming regions that radiatively
pump the molecules to the observed excitation state is a potential
appeal of this model, where stratification of the density profile and
turbulent properties of the cloud may account for the observed
variation in $b$ value with $J$ level. However, if the absorber
originates in the ISM of its host galaxy, the large velocity spread
among associated metal lines ($\Delta\,\varv \approx \pm 150$
km\,s$^{-1}$) suggests a significant circular velocity; thus, the DLA
is unlikely to be associated with the bound ISM of a low-mass
dwarf. Additionally, an origin in the ISM or CGM of a star-forming
galaxy is again in tension with the observed Fe enhancement of the
absorber.

We thus favor a model in which the H$_{2}$-bearing DLA is associated
with the outskirts of the massive, early-type galaxy at $d = 41$ pkpc,
as this model most fully reproduces the observed properties of the
absorber. The quiescent nature of this system is a natural explanation
for the apparent Fe enhancement of the absorber, a property that is
challenging to explain if the gas is associated with a star-forming
massive or dwarf galaxy. This system is only one of two known
H$_{2}$-bearing DLAs likely associated with a massive, quiescent
galaxy \citep{Zahedy2020}. QSO absorption-line spectroscopy thus
provides a powerful perspective on diffuse molecular gas properties in
the vicinity of early-type galaxies, offering an additional probe of
the cold gas reservoirs that persist in some systems despite the
quiescence of their host galaxies.

\section{Conclusion}
\label{sec:conc}

We report one of two known H$_{2}$-bearing DLAs likely associated with
a massive, early-type galaxy. Detected serendipitously in Lyman and
Werner band absorption at $z = 0.576$ towards J$0111\!-\!0316$ in the
CUBS program, this system highlights the importance of comprehensive
galaxy redshift survey data to probe the diversity of environments in
which a molecular medium is found.

This damped absorber (log[$N$(\ion{H}{1})/cm$^{-2}$] $= 20.1 \pm 0.2$)
displays complex kinematic structure with at least 10 velocity
components spread over $-130\ \text{km\,s$^{-1}$} \lesssim
\Delta\,\varv \lesssim 90$ km\,s$^{-1}$. Two of these components,
separated by $\Delta\,\varv \approx 60$ km\,s$^{-1}$, show Lyman and
Werner band absorption that implies a total H$_{2}$ column density of
log[$N$(H$_{2}$)/cm$^{-2}$] $= 18.97\substack{+0.05 \\ -0.06}$, or a
total H$_{2}$ fraction of $f_{\text{H}_{2}} = 0.11\substack{+0.06
  \\ -0.04}$. The excitation state of the molecular hydrogen and the
observed rise in $b$ value with $J$ level may be explained by
radiative pumping in a medium with stratification in turbulent
properties if the radiation field is elevated above that seen in the
solar neighborhood by one to two orders of magnitude.

A galaxy redshift survey reveals nine galaxies within $d = 600$ pkpc
and $\Delta\,\varv_{g} = 300$ km\,s$^{-1}$ of the absorber; the five
galaxies at the smallest impact parameter may form a bound group. The
galaxy at the closest projected distance to the QSO sightline is a
massive, early-type galaxy at $d = 41$ pkpc. An association betweeen
the absorber and the outer ISM or CGM of this galaxy is supported by
the chemical abundance pattern of the former, which suggests Fe
enhancement consistent with preferential enrichment by Type Ia SNe. We
discuss possible origins for the H$_{2}$-bearing gas in the
outskirts of this galaxy, including interactions with companion
galaxies and a past epoch of AGN activity. Although we cannot rule out
the presence of an undetected, intervening galaxy lost in the glare of
the QSO light, the lack of [\ion{O}{3}]\,$\lambda$5008
emission in the quasar spectrum at the absorber redshift places a
3$\sigma$ upper limit on the star formation rate of $\text{SFR} < 0.1$
$\text{M}_{\odot}$ yr$^{-1}$.

Current facilities, including ALMA, have detected dense molecular gas
in the central regions of galaxies with associated DLAs at
intermediate redshift \citep[e.g.,][]{Kanekar2018} and have the
ability to place sensitive limits on the diffuse molecular gas surface
density at the location of the QSO. Follow-up CO emission-line
observations probing the molecular gas reservoirs associated with the
early-type galaxy and its companions, including any undetected
galaxies, will help to more definitively determine the origin of the
molecular absorber.

At intermediate and high redshifts, QSO absorption-line spectroscopy
remains the only viable method to directly detect molecular gas at the
low and moderate column densities that may be characteristic of
circumgalactic environments (log$[N(\text{H}_{2})/\text{cm}^{-2}]
\lesssim 19$). Pairing such observations with comprehensive galaxy
survey data thus remains a critical avenue for gaining further clarity
about the diverse range of galactic evolutionary states and
environments that host dense and potentially star-forming gas.

\acknowledgments

We thank the referees for helpful comments, and Sergei
Balashev, Martin Bureau, Jay Gallagher, and Evan Kirby for useful
discussions. We are grateful to Dan Kelson for lending his expertise
on reducing the galaxy survey data from the Magellan Telescopes. EB,
HWC, and MCC acknowledge partial support from HST-GO-15163.001A and
NSF AST-1715692 grants. TC and GCR acknowledge support from
HST-GO-15163.015A. SDJ acknowledges support from a NASA Hubble
Fellowship (HST-HF2-51375.001-A). SC gratefully acknowledges support
from Swiss National Science Foundation grants PP00P2\_163824 and
PP00P2\_190092. KLC acknowledges partial support from NSF
AST-1615296. CAFG was supported by NSF through grants
AST-1517491, AST-1715216, and CAREER award AST-1652522, by NASA
through grant 17-ATP17-0067, by STScI through grants HST-GO-14681.011,
HST-GO-14268.022-A, and HST-AR-14293.001-A, and by a Cottrell Scholar
Award from the Research Corporation for Science Advancement. SL was
funded by project FONDECYT 1191232. This research was supported by the
Munich Institute for Astro- and Particle Physics (MIAPP) which is
funded by the Deutsche Forschungsgemeinschaft (DFG, German Research
Foundation) under Germany's Excellence Strategy – EXC-2094 –
390783311. This work is based on observations made with ESO Telescopes
at the Paranal Observatory under programme ID 0104.A-0147(A),
observations made with the 6.5m Magellan Telescopes located at Las
Campanas Observatory, and spectroscopic data gathered under the
HST-GO-15163.01A program using the NASA/ESA Hubble Space Telescope
operated by the Space Telescope Science Institute and the Association
of Universities for Research in Astronomy, Inc., under NASA contract
NAS 5-26555. This research has made use of NASA’s Astrophysics Data
System and the NASA/IPAC Extragalactic Database (NED) which is
operated by the Jet Propulsion Laboratory, California Institute of
Technology, under contract with the National Aeronautics and Space
Administration.

\software{\texttt{bagpipes} \citep{Carnall2018}, \texttt{CalCOS},
  \texttt{CarPy} \citep{Kelson2003}, \texttt{CLOUDY}
  \citep{Ferland2013}}





\end{document}